\pdfoutput=1

\documentclass[11pt,twoside,a4paper,cmspaper,final,collab]{cms-tdr}
\def\svnVersion{177011}\def\cmsCernNo{2012-156}\def\cmsCernDate{\today}\def\cmsMessage{Submitted to the Journal of High Energy Physics}

\begin{document}\cmsNoteHeader{SUS-11-011}

\hyphenation{had-ron-i-za-tion}
\hyphenation{cal-or-i-me-ter}
\hyphenation{de-vices}
\newcommand{\mll}{\ensuremath{m_{\ell\ell}}}

\newcommand{\eepm}{\Pep\Pem}
\newcommand{\mmpm}{\Pgmp\Pgmm}
\newcommand{\empm}{\ensuremath{\Pe^\pm \Pgm^\mp}}
\newcommand{\ase}[2]{\ensuremath{_{~- #1}^{~+ #2}}}
\newcommand{\met}{\ensuremath{\mspace{3mu}/\mspace{-12.0mu}E_{T}}}
\providecommand{\PYTHIA} {{\textsc{pythia}}\xspace}
\providecommand{\MADGRAPH} {\textsc{MadGraph}\xspace}

\newcommand{\tthad}{\ensuremath{\ttbar\to\ell^{\pm}+\text{jets}/\text{hadrons}}}
\newcommand{\lumifinal}{4.98\fbinv}
\newcommand{\ttll}{\ensuremath{\ttbar\to\ell^+\ell^-}}
\newcommand{\tttau}{\ensuremath{\ttbar\to\ell^{\pm}\Pgt^{\mp}/\Pgt^+\Pgt^-}}
\newcommand{\ttfake}{\ensuremath{\ttbar\to\text{fake}}}
\newcommand{\wjets}{\ensuremath{\PW+\text{jets}}}
\newcommand{\DY}{DY}
\newcommand{\WW}{\ensuremath{\PWp\PWm}}
\newcommand{\WZ}{\ensuremath{\PW^{\pm}\PZ}}
\newcommand{\ZZ}{\ensureamth{\PZ\PZ}}
\newcommand{\Ht}{\ensuremath{H_\mathrm{T}}}
\newcommand{\ptll} {\ensuremath{p_\mathrm{T}(\ell\ell)}}
\newcommand{\ptht} {\ensuremath{p_\mathrm{T}(\ell\ell/\sqrt{H_\mathrm{T}}}}
\newcommand{\cls}  {\ensuremath{\mathrm{CL_S}}}
\newcommand{\tauh}{\ensuremath{\Pgt_\mathrm{h}}}
\newcommand{\CTEQ} {{CTEQ}\xspace}
\providecommand{\re}{\ensuremath{\cmsSymbolFace{e}}} % base of natural logs

\newlength\cmsFigWidth
\ifthenelse{\boolean{cms@external}}{\setlength\cmsFigWidth{0.95\columnwidth}}{\setlength\cmsFigWidth{0.48\textwidth}}
\ifthenelse{\boolean{cms@external}}{\providecommand{\cmsLeft}{top}}{\providecommand{\cmsLeft}{left}}
\ifthenelse{\boolean{cms@external}}{\providecommand{\cmsRight}{bottom}}{\providecommand{\cmsRight}{right}}

\RCS$Revision: 157412 $
\RCS$HeadURL: svn+ssh://svn.cern.ch/reps/tdr2/papers/SUS-11-011/trunk/SUS-11-011.tex $
\RCS$Id: SUS-11-011.tex 157412 2012-11-12 02:12:38Z alverson $
\cmsNoteHeader{SUS-11-011} % This is over-written in the CMS environment: useful as preprint no. for export versions
\title{Search for new physics in events with opposite-sign leptons, jets, and missing transverse energy in pp collisions at $\sqrt{s}=7$\TeV}

\date{\today}

\abstract{
A search  is presented for physics  beyond the standard  model (BSM) in final states with a pair of opposite-sign
isolated leptons accompanied by jets and missing  transverse energy. The search uses LHC  data recorded at a center-of-mass energy
$\sqrt{s}=7$\TeV with  the CMS  detector, corresponding  to an integrated luminosity of approximately 5\fbinv.
Two complementary search strategies are employed.
The first probes models with a specific dilepton production mechanism
that leads to a characteristic kinematic edge in
the dilepton mass distribution. The second strategy probes models of dilepton production
with heavy, colored objects that decay to final states including invisible particles,
leading to very large hadronic activity and missing
transverse energy. No evidence for an event yield in excess of the standard model expectations is  found.
Upper  limits on  the  BSM contributions to the  signal regions are deduced from  the results, which are
used to exclude a region  of  the parameter space of the constrained minimal supersymmetric
extension of the standard model. Additional information
related to detector efficiencies and response is provided to allow testing specific models of BSM physics not
considered in this paper.
}

\hypersetup{%
pdfauthor={CMS Collaboration},%
pdftitle={Search for new physics in events with opposite-sign leptons, jets, and missing transverse energy in pp collisions at sqrt(s) = 7 TeV},%
pdfsubject={CMS},%
pdfkeywords={CMS, physics, supersymmetry}}

\maketitle %maketitle comes after all the front information has been supplied

\section{Introduction}
\label{sec:intro}

In this paper we describe a search for physics beyond the standard model (BSM)
in events containing a pair of opposite-sign leptons, jets, and missing transverse energy (\MET),
in a sample of proton-proton collisions at a center-of-mass energy of 7\TeV.
The data sample was collected with the Compact Muon Solenoid (CMS) detector~\cite{CMS:2008zzk} at
the Large Hadron Collider (LHC) in 2011
and corresponds to an integrated luminosity of 4.98\fbinv.
This is an update and extension
of a previous analysis performed with a data sample of 34\pbinv
collected in 2010~\cite{Chatrchyan:2011bz}.

The BSM signature in this search is motivated  by three general  considerations.
First, new particles predicted by BSM
physics scenarios are expected to be heavy in most cases, since they have so far eluded detection.
Second, BSM physics  signals may be produced with large cross section via the strong interaction,
resulting  in  significant hadronic  activity.
Third, astrophysical evidence for
dark matter  suggests~\cite{DM1,DM2,DM3,DM4}
that the mass  of weakly-interacting
massive particles is of the  order of the electroweak symmetry breaking
scale. Such particles, if produced in proton-proton collisions, could escape detection and give rise to
an apparent imbalance in the  event transverse energy. The analysis therefore focuses  on the
region  of  high \MET. An  example of  a  specific  BSM scenario  is
provided by R-parity conserving  supersymmetric (SUSY) models, in which
the colored squarks and gluinos  are  pair-produced  and  subsequently  undergo
cascade       decays,      producing  jets      and
leptons~\cite{Martin:1997ns,Wess:1974tw}.
These cascade decays may terminate in the production of the lightest SUSY particle (LSP),
often the lightest neutralino, which escapes detection and results in large \MET.
This LSP is a candidate for a dark matter weakly-interacting massive particle.
Another BSM scenario which may lead to similar signatures is the
model of universal extra dimensions (UED)~\cite{UEDColl}.

The results reported in this paper are part of a broad program of BSM searches
in events with jets and \MET, classified by the number and
type of leptons in the final state.
Here we describe a search for events containing an opposite-sign isolated lepton pair in addition to jets and \MET.
We reconstruct electrons and muons, which provide a clean signature with low background.
In addition, we reconstruct $\tau$ leptons in their hadronic decay modes to improve the sensitivity to models with
enhanced coupling to third generation particles.
Complementary CMS searches with different final states have
already been reported, for example in Refs.~\cite{ref:RA1,ref:SS}.
Results from the ATLAS collaboration in this final state using approximately 1--2\fbinv have been reported in Refs.~\cite{Aad:2011cwa,ATLAS:2012ag}.

The analysis strategy is as follows. In order to select dilepton events, we
use a preselection based
on that of the CMS top quark pair (\ttbar) cross section measurement in the dilepton channel~\cite{ref:top};
the details of this preselection are presented in Section~\ref{sec:eventSel}.
Reasonable agreement is found between the observed yields in data and the predictions from
standard model (SM) Monte Carlo (MC) simulation.
Two complementary search strategies are pursued, which are optimized for different experimental signatures.
The first strategy is a search for a kinematic edge~\cite{edge} in the dilepton ($\Pe\Pe$, $\Pgm\Pgm$) mass distribution.
This is a characteristic
feature of SUSY models in which the same-flavor opposite-sign leptons are produced via the decay
$\widetilde{\chi}_2^0 \to \ell\widetilde{\ell} \to \widetilde{\chi}_1^0 \ell^+\ell^-$, where $\widetilde{\chi}_2^0$
is the next-to-lightest neutralino,  $\widetilde{\chi}_1^0$ is the lightest neutralino, and $\widetilde{\ell}$ is a slepton.
The second strategy is a search for an excess of events with dileptons accompanied by very large hadronic activity and \MET.
We perform counting experiments in four signal regions with requirements on these quantities
to suppress the \ttbar\ background, and compare the observed yields
with the predictions from  a background estimation technique based on data control samples,
as well as with SM and BSM MC expectations.
These two search approaches are complementary, since the dilepton mass edge search is sensitive
to new physics models that have lower \MET and hadronic energy, while the counting experiments
do not assume a specific dilepton production mechanism  and are also sensitive to
BSM scenarios that produce lepton pairs with uncorrelated flavor.

No specific BSM physics scenario, e.g. a particular SUSY model, has been used to optimize the search regions.
In order to illustrate the sensitivity of the search, a simplified and practical model of
SUSY breaking, the constrained minimal supersymmetric
extension of the standard model (CMSSM)~\cite{CMSSM,CMSSM2} is used. The CMSSM is described by
five parameters: the universal scalar and gaugino mass parameters ($m_0$ and $m_{1/2}$, respectively),
the universal trilinear soft SUSY breaking parameter $A_0$, the
ratio of the vacuum expectation values of the two Higgs doublets ($\tan\beta$), and the sign of the
Higgs mixing parameter $\mu$.
Throughout the paper, four CMSSM parameter sets, referred
to as LM1, LM3, LM6, and LM13~\cite{PTDR2}, are used to illustrate possible CMSSM yields. The parameter
values defining LM1 (LM3, LM6, LM13) are $m_0 = 60~(330,85,270)$\GeV,
$m_{1/2} = 250~(240, 400, 218)$\GeV, $\tan\beta = 10~(20, 10,40)$, $A_0 = 0~(0,0,-553)$\GeV;
all four parameter sets have $\mu > 0$.
These four scenarios are beyond the exclusion reach of previous searches performed at the Tevatron and LEP,
and are chosen here because they produce events containing opposite-sign leptons and may lead to a kinematic
edge in the dilepton mass distribution.
These four scenarios serve as common benchmarks to facilitate comparisons of sensitivity among different analyses.

\section{The CMS detector}

The central feature of the CMS detector is a superconducting
solenoid, 13\unit{m} in length and 6\unit{m} in diameter, which provides
an axial magnetic field of 3.8\unit{T}. Within the field volume are
several particle detection systems.
Charged particle
trajectories are measured by silicon pixel and silicon strip trackers
covering $|\eta| < 2.5$ in pseudorapidity, where $\eta = -\ln [\tan \theta/2]$ with $\theta$ the
polar angle of the particle trajectory with respect to
the counterclockwise proton beam direction. A crystal electromagnetic calorimeter
and a brass/scintillator hadron calorimeter surround the
tracking volume, providing energy  measurements of electrons, photons and
hadronic jets. Muons are identified and measured in gas-ionization detectors embedded in
the steel return yoke outside the solenoid. The detector is nearly
hermetic, allowing energy balance measurements in the plane
transverse to the beam direction.
The first level of the CMS trigger system, composed of custom hardware processors, uses information from the calorimeters and muon detectors to select, in less than 1\mus, the most interesting events. The High Level Trigger processor farm further decreases the event rate from around 100\unit{kHz} to around 300\unit{Hz}, before data storage.
Event reconstruction is performed with the particle-flow (PF)
algorithm~\cite{CMS-PAS-PFT-10-002},
which is used to form a mutually exclusive collection
of reconstructed particles (muons, electrons, photons, charged
and neutral hadrons) by combining tracks and calorimeter clusters.
A more detailed description of the CMS detector can be found
elsewhere~\cite{CMS:2008zzk}.

\section{Event Selection}
\label{sec:eventSel}

The following samples of  simulated events are used to  guide the  design of  the analysis.
These      events     are      generated     with      either
\PYTHIA6.4.22~\cite{Pythia},  \MADGRAPH4.4.12~\cite{Madgraph}, or {\sc powheg}~\cite{POWHEG} MC event
generators using the \CTEQ6.6 parton density functions~\cite{cteq66}.
The \ttbar, $\PW+\text{jets}$, and $\mathrm{VV}$ ($\mathrm{V}=\PW,\Z$) samples are generated with \MADGRAPH,
with parton showering simulated by \PYTHIA using the Z2 tune~\cite{Chatrchyan:2011id}.
The single-top samples are generated with \POWHEG.
The Drell--Yan (DY) sample is generated using a mixture of \MADGRAPH\ (for events with dilepton invariant mass above 50\GeV)
and \PYTHIA (for events with dilepton invariant mass in the range 10--50\GeV), and includes decays to the $\Pgt\Pgt$ final state.
The signal events are simulated using \PYTHIA.
The detector response in these samples is then  simulated  with   a   \GEANTfour
model~\cite{Geant} of the CMS  detector. The MC events are reconstructed and
analyzed with the same software as is used to process collision data.
Due to the varying instantaneous LHC luminosity, the mean
number of interactions in a single beam crossing increased over the course of the data-taking period to a maximum of about 15.
In the MC simulation, multiple proton-proton interactions are simulated by \PYTHIA and superimposed on the hard collision,
and the simulated samples are reweighted to describe the distribution
of reconstructed primary vertices in data~\cite{Khachatryan:2010pw}.
The simulated sample yields are normalized to an integrated luminosity of ~\lumifinal\ using
next-to-leading order (NLO) cross sections.

Events in data are selected with a set of $\Pe\Pe$, $\Pe\Pgm$, $\Pgm\Pgm$, $\Pe\Pgt$,
and $\mu\tau$ double-lepton triggers.
Since the online reconstruction of hadronic-$\Pgt$ decays (\tauh) is difficult,
\tauh\ triggers are intrinsically prone to high rates.
Therefore,  for the analysis with two \tauh\ only, we use specialized triggers
that rely on significant hadronic activity $\HT$, quantified by the scalar sum of online jet transverse energies with \pt\ $>$ 40 GeV,
and $\MET$ as well as the presence of two \tauh.
The  efficiencies for events containing two
leptons passing the analysis selection to  pass at  least  one of  these
triggers are measured to be approximately
$1.00^{+0.00}_{-0.02}$, $0.95\pm0.02$, $0.90\pm0.02$, $0.80\pm0.05$, $0.80\pm0.05$ and $0.90\pm0.05$
for $\Pe\Pe$, $\Pe\mu$, $\mu\mu$, $\Pe\tauh$, $\mu\tauh$ and $\tauh\tauh$ triggers, respectively.
In the following, the simulated sample yields for the light lepton channels are weighted by these trigger efficiencies.
For the \tauh\ channels the trigger simulation is applied to the MC simulation and then a correction is applied
based on the measured data and MC efficiencies for these triggers.

Because leptons produced in the decays of low-mass particles, such as hadrons containing b
and c quarks, are nearly always inside jets, they can be suppressed by requiring the leptons
to be isolated in space from other particles that carry a substantial amount of transverse momentum.
The details   of   the  lepton   isolation   measurement   are  given   in Ref.~\cite{ref:top}.
In  brief,   a cone is constructed     of  size
$\Delta{}R\equiv\sqrt{(\Delta\eta)^2+(\Delta\phi)^2}=0.3$  around  the
lepton  momentum  direction. The  lepton  relative  isolation is  then
quantified  by  summing the  transverse  energy  (as  measured in  the
calorimeters) and the transverse  momentum (as measured in the silicon
tracker) of  all objects  within this cone,  excluding the  lepton, and
dividing by  the lepton transverse momentum. The resulting quantity
is required to be  less than 0.15, rejecting
the large background arising from QCD production of jets.

The \tauh\ decays are reconstructed with
the PF algorithm and identified with the
hadrons-plus-strips (HPS) algorithm, which considers candidates with one or
three charged pions and up to two neutral pions~\cite{TauPAS}.
As part of the \tauh\ identification procedure, loose isolation is
applied for the $\tauh$ final states.
Isolated electrons and muons can be misidentified as \tauh\ candidates.
For this reason \tauh\ candidates are
required to fail electron selections and
not to match a muon signature in the muon system.

Events with two opposite-sign isolated leptons are selected. At least one of the leptons must
have  $\pt >  20$\GeV, both must  have $\pt  > 10$\GeV,  and the
electrons (muons) must have $|\eta| < 2.5$ ($|\eta| < 2.4$). Electrons in the range $1.44 < |\eta| < 1.57$
are excluded. In events containing
a $\tauh$ candidate, both leptons must satisfy $\pt >  20$\GeV and $|\eta| < 2.1$, where the acceptance
requirement is tightened so that the \tauh\ decay products are contained in the tracking detector
in a manner that is consistent with the requirements of the triggers used for these events.
In events with more than one opposite-sign pair that satisfy the selection
requirements, the two oppositely-signed leptons with highest \pt are chosen.
Events with an $\Pe\Pe$  or $\Pgm\Pgm$ pair
with  invariant mass of the dilepton system between 76\GeV  and  106\GeV or  below
12\GeV are removed, in  order      to     suppress
$\cPZ/\gamma^{*}\to\ell\ell$  events,  as  well  as   low-mass  dilepton
resonances. Events containing two electrons, two muons, or an electron and a muon are referred to as the ``light lepton channels,'' while
events with at least one \tauh\ are referred to as ``hadronic-$\tau$ channels.''

The PF objects are clustered to form jets using the anti-$k_{\mathrm{T}}$ clustering algorithm~\cite{antikt} with the distance parameter of 0.5.
We apply \pt- and $\eta$-dependent corrections to account for residual effects of nonuniform detector response,
and impose quality criteria to reject jets that are consistent with anomalous detector noise.
We require the presence of at least two jets with transverse momentum of $\pt > 30$\GeV and $|\eta| < 3.0$,
separated  by $\Delta  R  >$  0.4 from  leptons  passing the  analysis selection. For each event the scalar
sum of transverse energies of selected jets $\HT$ must exceed $100$\GeV.
The \MET is defined as the magnitude of the vector sum of the transverse momenta of all PF objects,
and we require $\MET>50$\GeV ($\MET>100$\GeV) in the light lepton (hadronic-$\Pgt$) channels.

\begin{table*}[t]
\begin{center}
\topcaption{\label{tab:cuts}
Summary of event preselection requirements applied in the light lepton channels, hadronic-$\Pgt$ channels, and the dilepton mass
edge search of Section~\ref{sec:fit}. The leading (trailing) lepton is the one with highest (second highest) \pt. The
requirements on jet multiplicity, scalar sum of jet transverse energies (\Ht), missing transverse energy (\MET), and dilepton
mass are also indicated.
}
\begin{tabular}{l|c|c|c}
\hline
\hline
Requirement & light leptons & hadronic-$\tau$ & edge search \\
\hline
leading lepton   & e or $\mu$, \pt $>$ 20\GeV & e, $\mu$, or $\tauh$, \pt $>$ 20\GeV & e or $\mu$, $\pt > 20\GeV$ \\
trailing lepton  & e or $\mu$, \pt $>$ 10\GeV & e, $\mu$, or $\tauh$, \pt $>$ 20\GeV & e or $\mu$, $\pt > 10\GeV$ \\
jet multiplicity & $n_\text{jets}\geq2$ & $n_\text{jets}\geq2$ & $n_\text{jets}\geq2$ \\
\Ht              & $\Ht> 100\GeV$ & $\Ht> 100\GeV$ & $\Ht> 300\GeV$ \\
\MET             & $\MET> 50\GeV$ & $\MET> 100\GeV$ & $\MET> 150\GeV$ \\
dilepton mass    & veto $76<m_{ee},m_{\mu\mu}<106\GeV$ & - & - \\
\hline
\hline
\end{tabular}
\end{center}
\end{table*}

The event preselection requirements are summarized in Table~\ref{tab:cuts}.
The data yields and corresponding MC predictions after this event preselection
are given in Table~\ref{tab:yields} (light leptons) and Table~\ref{tab:yieldsTau} (hadronic-$\tau$).
For the light lepton channels, the normalization of the simulated yields has been scaled based on studies
of $\Z\to\ell\ell$ in data and in MC simulation, to account for effects of lepton selection and trigger efficiency
and to match the integrated luminosity.
As expected, the MC simulation predicts that the sample passing the preselection is dominated by lepton pair final states from \ttbar decays (dilepton \ttbar).
The data yield is in good agreement with the prediction, within the systematic uncertainties of the integrated luminosity
(2.2\%) and \ttbar cross section determination (12\%)~\cite{top1,top2,top3}.
The yields for the LM1, LM3, LM6, and LM13 benchmark scenarios are also quoted.

\begin{table*}[htbp]
\begin{center}
\footnotesize
\small
\topcaption{\label{tab:yields} Data yields and MC predictions in the light lepton channels after preselection, using the quoted NLO production cross sections $\sigma$.
The \ttll\ contribution corresponds  to dilepton \ttbar with no $\PW\to\Pgt$ decays, \tttau refers to dilepton \ttbar with at least one $\PW\to\tau$ decay,
and \tthad\ includes all other \ttbar decay modes.
The quoted cross sections for these processes include the relevant branching fractions.
The LM points are benchmark SUSY scenarios, which are defined in the text.
The MC uncertainties include the statistical component, the uncertainty in the integrated luminosity,
and the dominant uncertainty from the \ttbar cross-section determination.
The data yield is in good agreement with the MC prediction, but the latter is not used explicitly in the search.
The difference between the $\Pe\Pe+\Pgm\Pgm$ versus $\Pe\Pgm$ yields is due to the
rejection of $\Pe\Pe$ and $\Pgm\Pgm$ events with an invariant mass consistent with that of the \Z boson.
}
\begin{tabular}{lr|cccc}

\hline
\hline
         Sample & $\sigma$ [pb]  &       $\Pe\Pe$ &        $\Pgm\Pgm$ &          $\Pe\Pgm$ &             Total \\

\hline
          \ttll &        7       &      1466 $\pm$ 179 &      1872 $\pm$ 228 &         4262 $\pm$ 520 &       7600 $\pm$ 927         \\

         \tttau &        9       &       303 $\pm$  37 &       398 $\pm$  49 &          889 $\pm$ 108 &       1589 $\pm$ 194         \\

         \tthad &      141       &        50 $\pm$ 6.2 &        15 $\pm$ 1.9 &           90 $\pm$  11 &        155 $\pm$  19         \\

DY$\to\ell\ell$ &     16677      &       193 $\pm$ 11  &       237 $\pm$  13 &          312 $\pm$  15 &        741 $\pm$ 26          \\
         \PW\PW &       43       &        55 $\pm$ 1.7 &        66 $\pm$ 1.9 &          151 $\pm$ 3.8 &        272 $\pm$ 6.5         \\
          \PW\Z &       18       &        13 $\pm$ 0.4 &        15 $\pm$ 0.4 &           25 $\pm$ 0.6 &         53 $\pm$ 1.3         \\
           \cPZ\cPZ &      5.9       &       2.6 $\pm$ 0.1 &       3.3 $\pm$ 0.1 &          3.3 $\pm$ 0.1 &        9.1 $\pm$ 0.3         \\
     Single top &      102       &        95 $\pm$ 3.1 &       120 $\pm$ 3.7 &          278 $\pm$ 7.3 &        492 $\pm$ 12          \\
         \wjets &    96648       &        47 $\pm$  11 &       9.8 $\pm$ 4.6 &           59 $\pm$  12 &        117 $\pm$ 16          \\
\hline
\hline
       Total MC &                &      2224 $\pm$ 224 &      2735 $\pm$ 281 &         6069 $\pm$ 643 &      11029 $\pm$ 1137       \\
           Data &                &                2333 &                2873 &                  6184 &              11390           \\
\hline
\hline
            LM1 &    6.8         &       272 $\pm$ 8.3 &      342 $\pm$ 9.7 &         166 $\pm$ 5.7 &        780 $\pm$ 20          \\
            LM3 &    4.9         &       107 $\pm$ 3.7 &      125 $\pm$ 4.1 &         181 $\pm$ 5.5 &        413 $\pm$ 11          \\
            LM6 &    0.4         &        20 $\pm$ 0.6 &       23 $\pm$ 0.7 &          26 $\pm$ 0.8 &         69 $\pm$ 1.7         \\
            LM13&    9.8         &       138 $\pm$ 6.6 &      157 $\pm$ 7.0 &         334 $\pm$  12 &        629 $\pm$ 19          \\
\hline

\hline
\end{tabular}
\end{center}
\end{table*}

\begin{table*}[hbtp]
\begin{center}
\footnotesize
\topcaption{\label{tab:yieldsTau} Data yields and MC predictions in hadronic-$\Pgt$ channels after preselection, using the quoted NLO production cross sections $\sigma$. Diboson backgrounds comprise \PW\PW, \PW\Z and $\cPZ\cPZ$ events. The sum of simulated events is also split into events with a generated \tauh\ (MC, genuine \tauh)
and events with a misidentified \tauh\ (MC, misidentified \tauh);
the two contributions are equally important.
The channel with two \tauh\ decays is not presented because the trigger
is not efficient in the preselection region, due to the large $\Ht$ requirement.
The uncertainty indicated represents both statistical and systematic components.
}
\begin{tabular}{lr|cc|c}
\hline
\hline
Sample &  $\sigma$ [pb] & $\Pe\tauh$ & $\mu\tauh$ & Total\\
\hline
DY$\to\ell\ell$             & 16677 & 51 $\pm$ 12 & 47 $\pm$ 11 & 98 $\pm$ 22\\
\ttbar                      & 157.5 & 165 $\pm$ 47 & 205 $\pm$ 58 & 370 $\pm$ 105\\
Diboson                     & 66.9  & 11 $\pm$ 2.0 & 10.8 $\pm$ 1.9 & 22 $\pm$ 3.6\\
Single top                  & 102  & 7.2 $\pm$ 2.6 & 8.1 $\pm$ 2.7 & 15 $\pm$ 4.8\\
\hline
$\sum\text{MC, genuine }\tauh$        &       & 146 $\pm$ 39 & 167 $\pm$ 44 & 313 $\pm$ 83\\
$\sum\text{MC, misidentified }\tauh$        &      & 89 $\pm$ 24 & 103 $\pm$ 27 & 191 $\pm$ 51\\
\hline
Total MC            &        & 235 $\pm$ 62 & 271 $\pm$ 72 & 505 $\pm$ 134\\
\hline
$\text{Data}$ & & $215$ & $302$ & $517$\\
\hline
\hline
LM1                         & 6.8  & 36 $\pm$ 6.7 & 46 $\pm$ 6.8 & 82 $\pm$ 9.8\\
LM3                        & 4.9    & 28 $\pm$ 6.0 & 18 $\pm$ 4.6 & 46 $\pm$ 7.6\\
LM6                         & 0.4    & 2.8 $\pm$ 1.1 & 4.2 $\pm$ 1.3 & 7.0 $\pm$ 1.7\\
LM13                        & 9.8    & 90 $\pm$ 11 & 118 $\pm$ 12 & 208 $\pm$ 16\\
\hline
\hline
\end{tabular}
\end{center}
\end{table*}

\section{Search for a Kinematic Edge}
\label{sec:fit}

We search for a kinematic edge (end-point) in the dilepton mass distribution for same-flavor (SF)
light-lepton events, i.e., $\Pe\Pe$ or $\Pgm\Pgm$ lepton pairs.
Such an edge is a characteristic feature of, for example, SUSY scenarios in which
the opposite-sign leptons are produced via the decay $\widetilde{\chi}_2^0 \to \ell\widetilde{\ell} \to \widetilde{\chi}_1^0 \ell^+\ell^-$. The model of UED
can lead to a similar signature with different intermediate particles.
In case of a discovery such a technique offers one of
the best possibilities for model-independent constraints
of the SUSY mass parameters~\cite{edge}.

In contrast, for the dominant background \ttbar\ as well as other
SM processes such as \PW\PW\ and $\mathrm{DY}\to\Pgt\Pgt$, the two lepton flavors are uncorrelated,
and the rates for SF and opposite-flavor (OF) $\Pe\Pgm$ lepton pairs are therefore the same.
Hence we can search for new physics in the SF final state and model the backgrounds
using events in the OF final state. Thus the \ttbar\ background shape is extracted from
events with OF lepton pairs, and a fit is performed
to the dilepton mass distribution in events with SF lepton pairs.

\begin{figure}[htbp]
\begin{center}
\includegraphics[width=\cmsFigWidth]{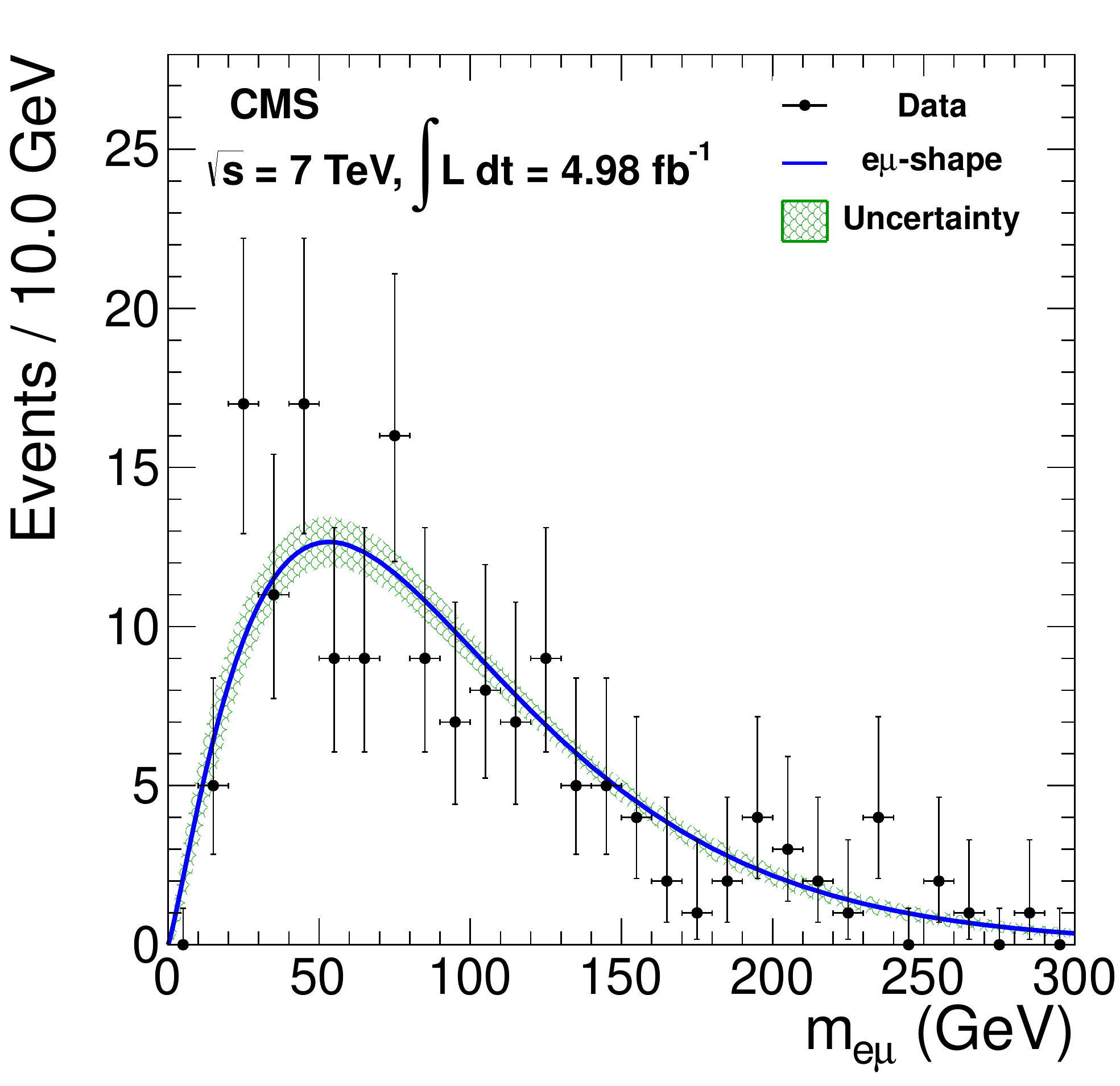}
\includegraphics[width=\cmsFigWidth]{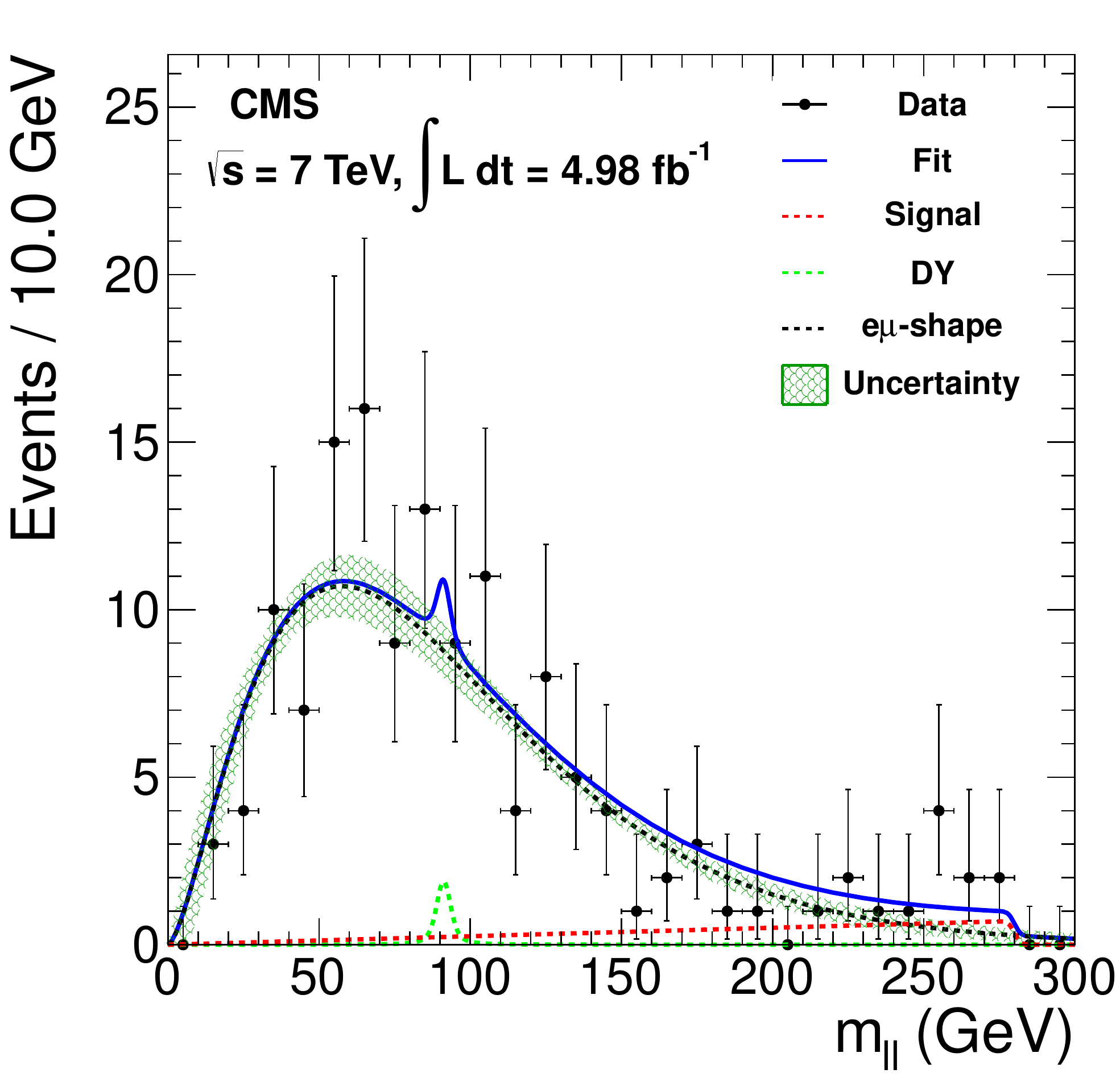}
\caption{\label{fig:dilmass}\protect
Distribution of events (black points) and the results of
the maximum likelihood fit (blue curve) to the dilepton mass distribution
for events containing $\Pe\Pgm$ lepton pairs (\cmsLeft), and $\Pe\Pe$ and $\Pgm\Pgm$ lepton pairs (\cmsRight) in the signal region
$\Ht> 300\GeV$ and $\MET> 150\GeV$, that suppresses DY contributions almost completely. The signal hypothesis for a value of the kinematic edge position $m_\text{max}=280$\GeV, corresponding to the largest local excess,  is displayed.
The shaded band represents the shape uncertainty of the background model.
}
\end{center}
\end{figure}

In order to be sensitive to BSM physics over the full dilepton mass spectrum,
events with a dilepton invariant mass $\mll$ consistent with
that of the \Z boson are not rejected.
This increases the DY contribution, which is compensated
by an increase in the $\MET > 150\GeV$ requirement (see Table~\ref{tab:cuts}).
We then proceed to search for a kinematic edge in the signal region defined as
$\Ht> 300\GeV$. The invariant mass distributions of SF and OF  lepton pairs
are in good agreement with each other (see Fig.~\ref{fig:dilmass}).
A fit is performed to the dilepton mass distribution with three candidate signal shapes,
over a range of values on the position of the kinematic edge.

The flavor-uncorrelated background, as a function of the invariant mass $\mll$, is parameterized as:
\begin{equation}
B(\mll) = \mll^a\,\re^{-b \mll},
\end{equation}
where $a\approx1.4$ describes the rising edge and $b\approx0.002$
dominates the long exponential tail on the right hand side of the
background shape; these parameters are extracted from the fit to data.

We parametrize the signal shape with an edge model for two subsequent two-body decays,
according to:
\begin{equation}
\label{eq:triangle}
S(\mll) = \frac{1}{\sqrt{2\pi}\sigma_{ll}}\int_0^{m_\text{max}} \rd y~y^\alpha \re^{-\frac{(\mll-y)^2}{2\sigma_{ll}^2}}.
\end{equation}

For $\alpha=1$ this function describes a triangle convoluted with a Gaussian, which accounts for detector resolution
effects. The resolution parameters for electrons $\sigma_{\Pe\Pe}$ and muons $\sigma_{\Pgm\Pgm}$ are constrained
with simulation.
The DY contribution, found to be negligible as seen in Fig.~\ref{fig:dilmass},
is modelled by a Breit-Wigner function with the mass and width parameters
fixed at the $\cPZ$ boson mass and width,
convoluted with a Gaussian function to account for the detector resolution.

We perform a simultaneous, extended, unbinned maximum
likelihood fit to the distribution of dilepton mass for events containing $\Pe\Pe$, $\mu\mu$
(signal, DY and background model),
and $\Pe\Pgm$ pairs (background model only).
The value of the kinematic edge position $m_\text{max}$ is varied, and the fit is performed for each value of this parameter.
The shape parameters of the flavor-uncorrelated background that are free in the fit
are assumed to be common in all categories,
and the yields of signal ($n_\mathrm{S}$), DY ($n_\mathrm{DY}$) and background ($n_\mathrm{B}$)
in these three categories are constrained using the ratio of muon to electron selection efficiencies
$R_{\Pgm\Pe} = 1.11 \pm 0.05$. This quantity is evaluated using studies of DY events in data and in MC simulation.

The fit is performed in the signal region $\Ht> 300\GeV$ and $\MET > 150\GeV$. The SF events
overlaid with the signal plus background fit, and the flavor-uncorrelated shape overlaid with OF events,
are shown in Fig.~\ref{fig:dilmass}.
The results of the fit are displayed for a value of the kinematic edge position $m_\text{max}=280\GeV$,
where the largest excess is observed. The local significance is $2.1\sigma$ including statistical and systematic
uncertainties. However,
a correction for the look-elsewhere effect~\cite{LEE} reduces the global significance to $0.7\sigma$.
The extracted signal yield including statistical uncertainty ($n_\mathrm{S} = 11_{- 5.7}^{+6.5}$)
at this point is consistent with the background-only hypothesis,
and we derive a 95\% confidence level upper limit of $n_\mathrm{S} < 23$~events for
this kinematic edge position.
No evidence for a kinematic edge feature is observed in the dilepton mass distribution.
\section{Counting Experiments}
\label{sec:datadriven}

We next proceed to search for an excess of events containing lepton pairs accompanied by large \MET\ and \Ht.
To look for possible BSM contributions, we define four signal regions that reject all but
$\sim$0.1\% of the dilepton \ttbar events, by adding the following requirements:

\begin{itemize}
\item high-\MET  signal region   :  $\MET> 275\GeV$, $\Ht> 300\GeV$,
\item high-\Ht\ signal region    :  $\MET> 200\GeV$, $\Ht>600\GeV$,
\item tight signal region :  $\MET> 275\GeV$, $\Ht > 600\GeV$,
\item low-\Ht\ signal region     :  $\MET> 275\GeV$, $125 < \Ht <300\GeV$.
\end{itemize}
The signal regions are indicated in Fig.~\ref{fig:met_ht}.
These signal regions are tighter than the one used
in Ref.~\cite{Chatrchyan:2011bz} since with the larger data sample the tighter signal regions allow us to explore
phase space farther from the core of the SM distributions. The observed and estimated yields in the high-\MET,
high-\Ht, and tight signal regions are used in the CMSSM exclusion limit in Section~\ref{sec:limit}.
The low-\Ht\ region has limited sensitivity to CMSSM models that tend to produce low-\pt\ leptons,
since the large \MET\ and low \Ht\ requirements lead to the requirement of large dilepton \pt.
However, the results of this region are included to extend the sensitivity to other models that produce high-\pt\ leptons.

\subsection{Light lepton channels}

The dominant background in the signal regions is dilepton \ttbar production.
This background is estimated using a technique
based on data control samples, henceforth referred to
as the dilepton transverse momentum ($\pt(\ell\ell)$) method. This method
is  based on the fact~\cite{ref:victory} that  in dilepton  \ttbar  events the
\pt\  distributions of  the charged  leptons (electrons and muons) and  neutrinos
are  related, since each lepton-neutrino pair is produced in the two-body decay of the $\PW$ boson.
This relation depends on the polarization  of the $\PW$ bosons,
which         is         well         understood        in         top quark
decays in the SM~\cite{Wpolarization,Wpolarization2,Wpolarization3},   and   can  therefore   be
reliably  accounted   for.
In dilepton \ttbar\ events, the values of \ptll\ and the transverse momentum of the dineutrino system ($\pt(\Pgn\Pgn)$)
are approximately uncorrelated on an event-by-event basis.
We thus  use   the  observed $\pt(\ell\ell)$ distribution to  model the $\pt(\Pgn\Pgn)$ distribution,
which is  identified with \MET. Thus, we predict the background in a signal region $S$
defined by requirements on \MET\ and \Ht\ using the yield in a region $S'$
defined by replacing the \MET\ requirement by the same requirement on $\pt(\ell\ell)$.

To suppress the DY contamination to the region $S'$, we increase the \MET\ requirement to $\MET> 75\GeV$ for SF events and subtract off the
small residual DY contribution using the $R_\text{out/in}$ technique~\cite{ref:top} based on control samples in data.
This technique derives, from the observed DY yield in the \cPZ\ mass region, the expected yield
in the complementary region using the ratio $R_\text{out/in}$ extracted from MC simulation.
Two corrections are applied to the resulting prediction, following the same procedure as in Ref.~\cite{Chatrchyan:2011bz}.
The first correction accounts for the fact that we apply minimum requirements to \MET in the preselection
but there is no corresponding requirement on \ptll. Since the \MET\ and \ptll\ are approximately uncorrelated in individual dilepton \ttbar
events, the application of the \MET\ requirement decreases the normalization of the \ptll\ spectrum without significantly altering the shape.
Hence, we apply correction factors $K$, which are extracted from data as
$K =1.6 \pm 0.1$, $1.6 \pm 0.4$, $1.6 \pm 0.4$, and $1.9\pm0.1$ for the high-\MET, high-\Ht, tight, and low-\Ht\ signal regions, respectively.
The uncertainty in $K$ is dominated by the statistical component.
The  second correction factor $K_C$ accounts for the $\PW$ polarization in \ttbar\ events, as well
as detector effects such as hadronic energy scale; this correction is extracted from MC and is
$K_C  = 1.6  \pm 0.5$, $1.4 \pm 0.2$, $1.7 \pm 0.4$, and $1.0 \pm 0.4$ for the four respective regions.
The uncertainty in $K_C$ is dominated by MC sample statistics and by the 7.5\% uncertainty in the hadronic energy scale in this analysis.

Backgrounds from DY are estimated from data with the $R_\text{out/in}$ technique,
which leads to an estimated DY contribution consistent with zero.
Backgrounds from processes with two vector bosons as well as
electroweak single top quark production
are negligible compared with those from dilepton \ttbar decays.

Backgrounds  in  which one  or  both  leptons  do not  originate  from
electroweak decays (misidentified leptons) are assessed  using the ``tight-to-loose'' (TL) ratio ($R_\mathrm{TL}$)
method of Ref.~\cite{ref:top}.
A misidentified lepton is  a lepton  candidate
originating from within a jet,  such as a lepton from semileptonic b
or  c decays,  a muon from a pion or kaon decay-in-flight, a  pion misidentified  as an
electron,  or an  unidentified  photon conversion.
The results of the tight-to-loose ratio method confirm the MC
expectation that the misidentified lepton contribution is small compared to the
dominant backgrounds. Estimates of  the
contributions to  the signal region  from QCD multijet events,  with two
misidentified leptons, and in $\PW+\text{jets}$, with one misidentified lepton
in  addition to  the lepton  from the  decay of  the $\PW$,  are derived
separately. The contributions are found to be less than $\sim$10\% of the total
background in the signal regions, which is comparable to the contribution in the
control regions used to estimate the background from the \ptll\ method.
We therefore assign an additional systematic uncertainty of 10\% on the background prediction from the \ptll\ method
due to misidentified leptons.

\begin{figure}[htbp]
\begin{center}
\includegraphics[width=\cmsFigWidth]{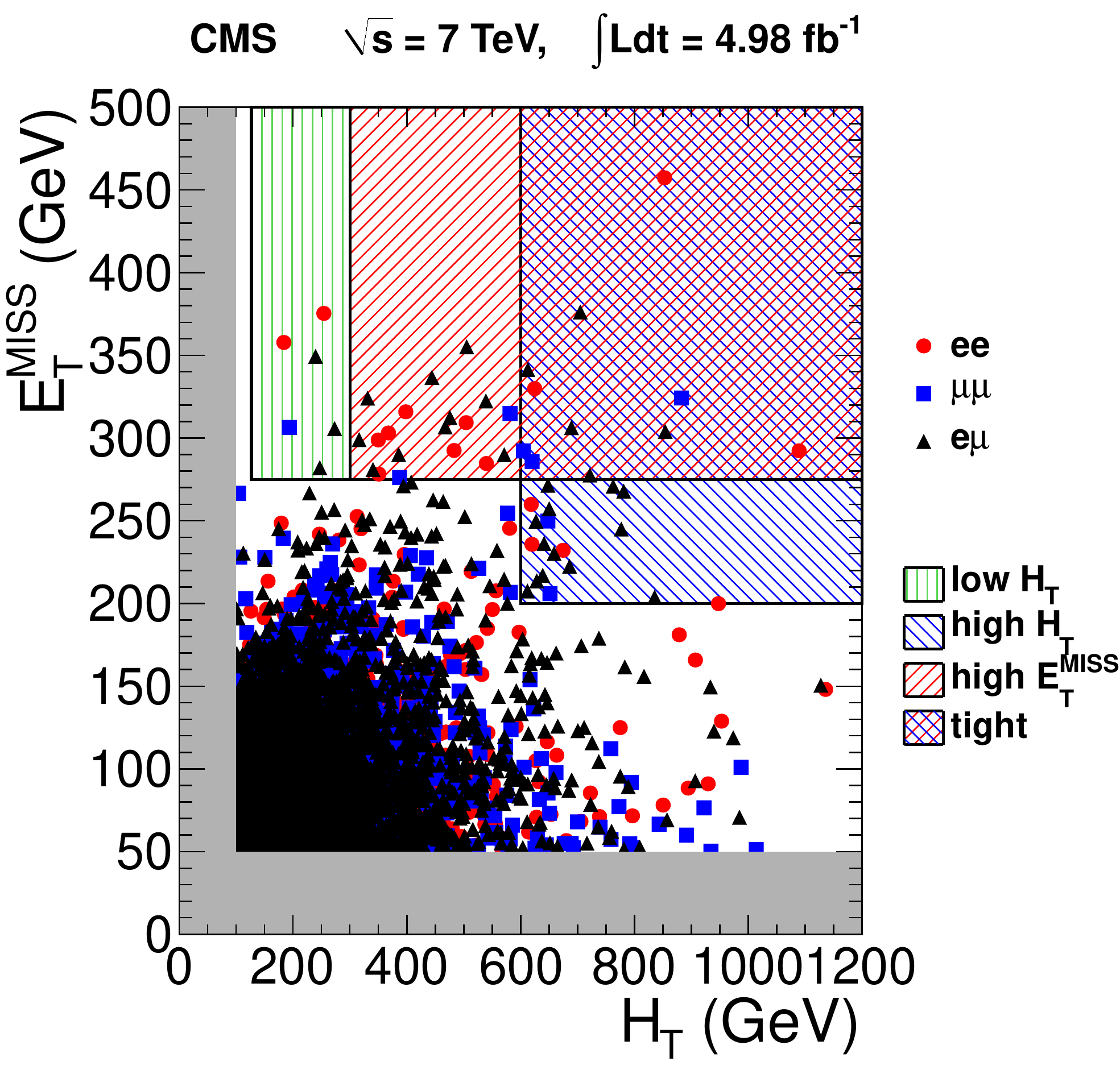}
\includegraphics[width=\cmsFigWidth]{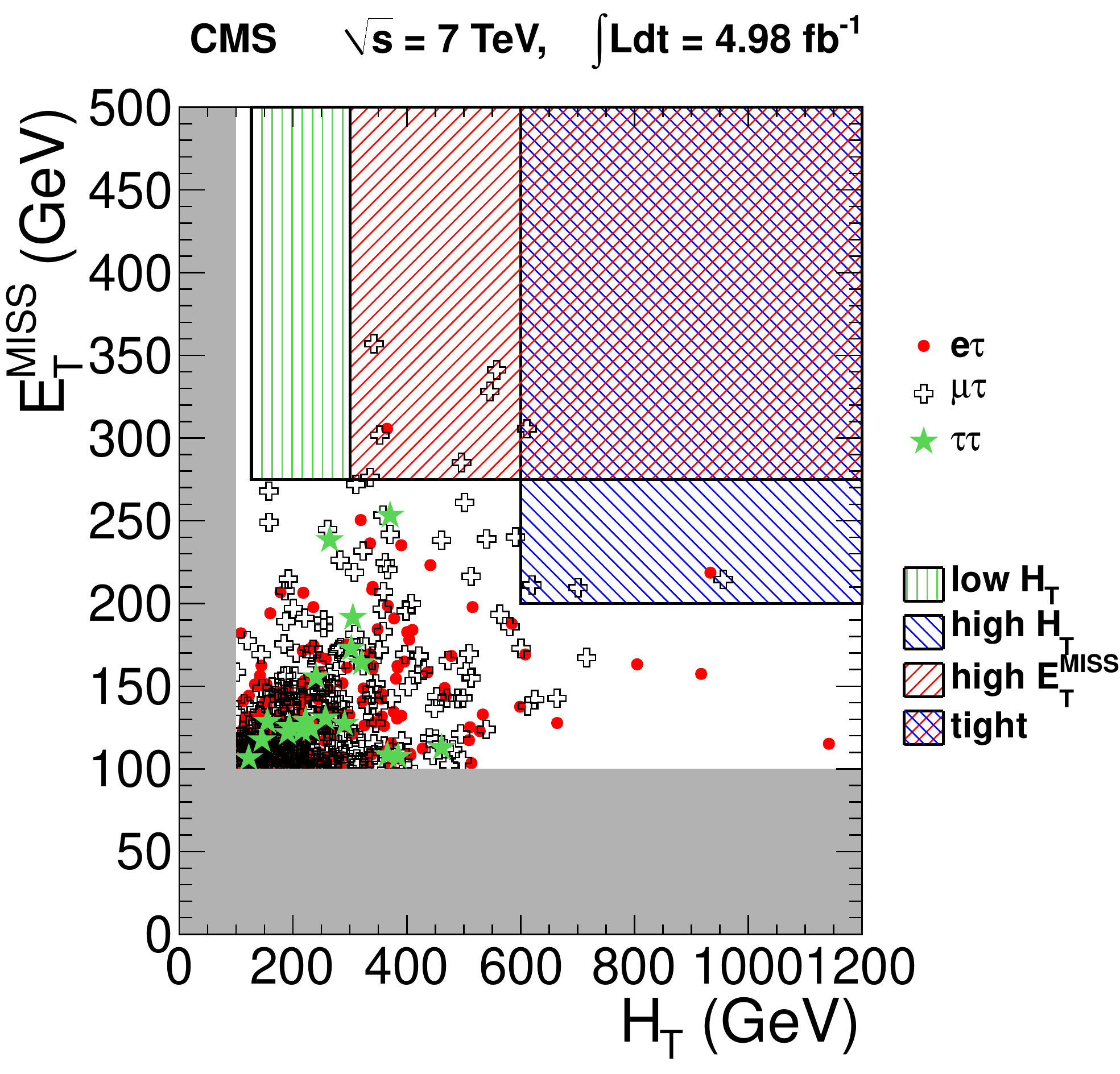}
\caption{\label{fig:met_ht}\protect Distributions of \MET\ vs.\ \HT\
for data in the light lepton channels (\cmsLeft) and hadronic-$\Pgt$ channels (\cmsRight).
The signal regions are indicated as hatched regions. The solid gray region is
excluded at the preselection level.
}
\end{center}
\end{figure}

\begin{figure}[htbp]
\begin{center}
\includegraphics[width=\cmsFigWidth]{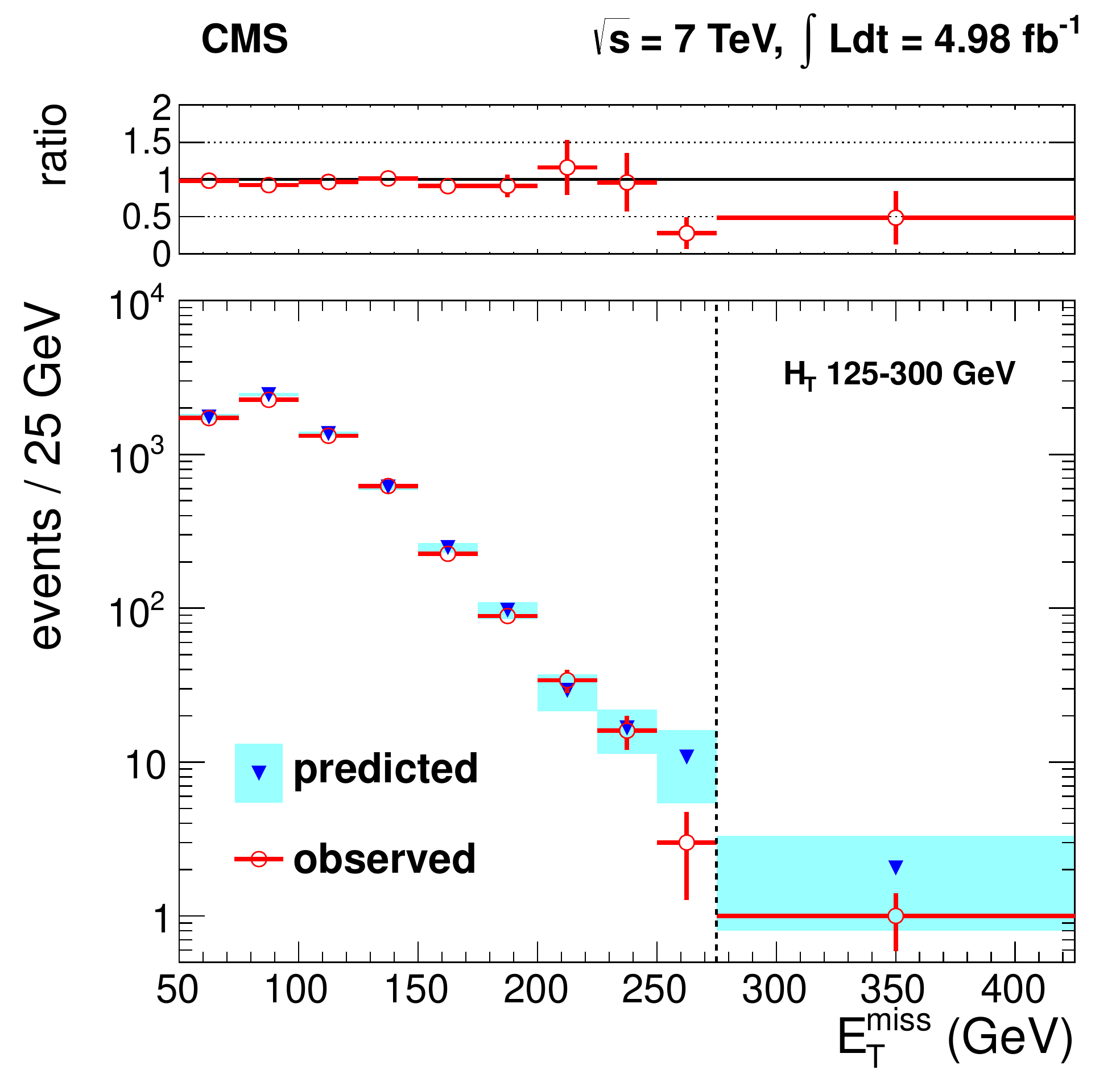}
\includegraphics[width=\cmsFigWidth]{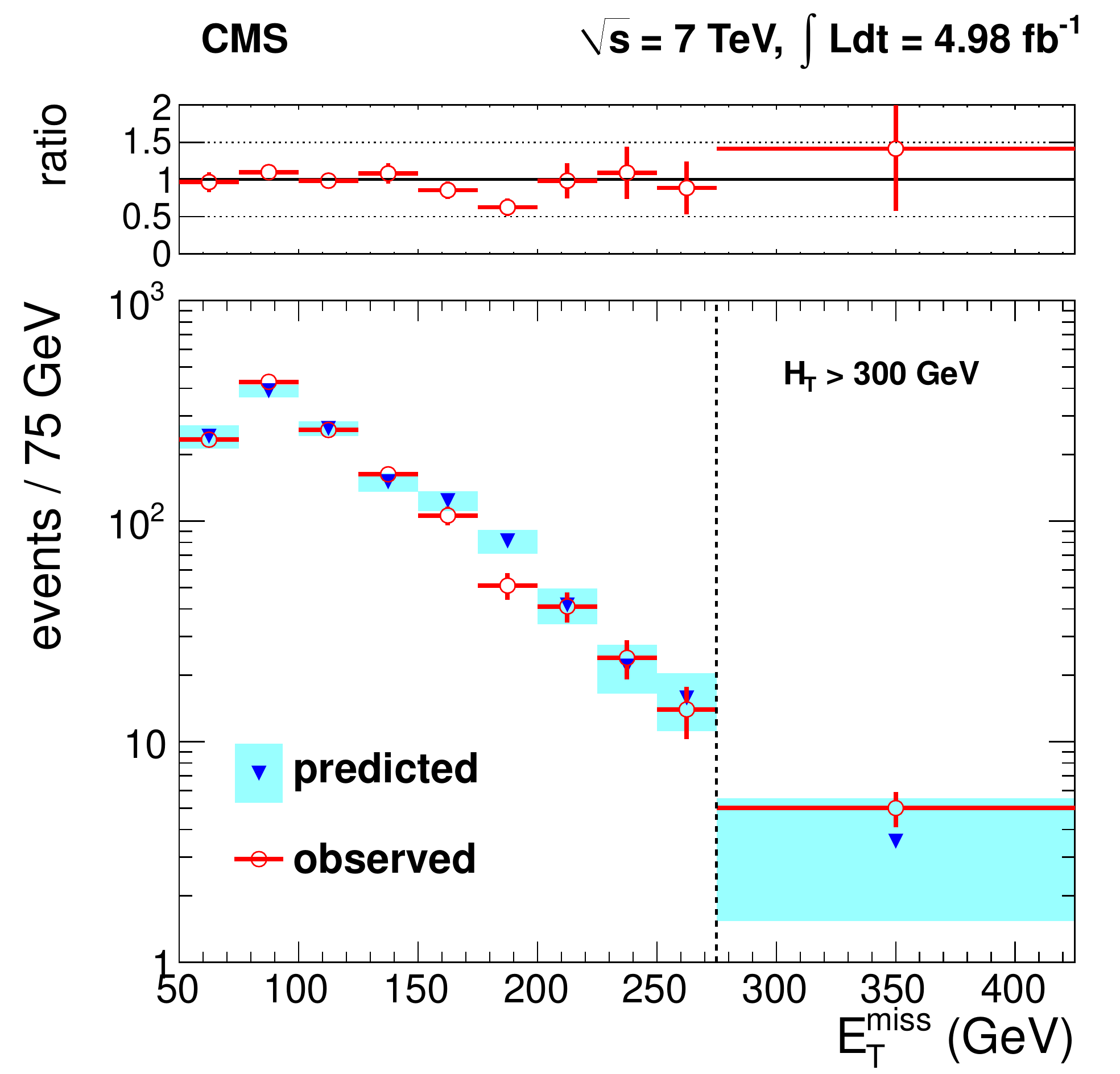}
\includegraphics[width=\cmsFigWidth]{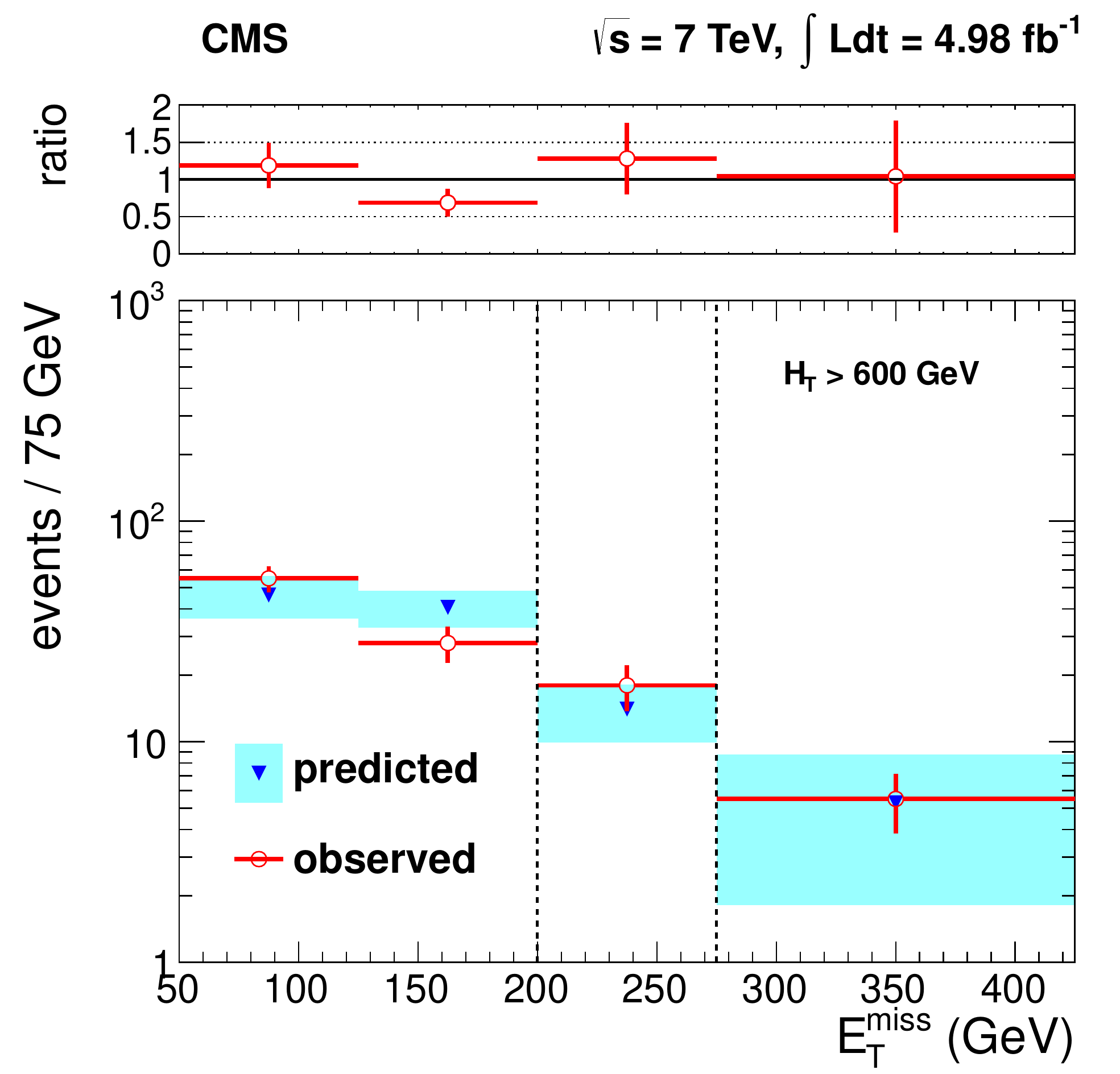}
\caption{\label{fig:victory}\protect
The observed \MET\ distributions (red points) and \MET\ distributions predicted by the \ptll\ method (blue points with shaded uncertainty bands) in data
for the region $125 < \Ht < 300\GeV$ (upper left), $\Ht> 300\GeV$ (upper right), and $\Ht> 600\GeV$ (bottom).
The uncertainty bands on the predicted \MET\ distribution are statistical, and also include systematic uncertainties for points in the signal regions,
to the right of the vertical dashed line. The ratio of data to predicted background is also included. The error bars include the full uncertainties
on the data and predicted background.
}
\end{center}
\end{figure}

As a validation of the $\pt(\ell\ell)$ method in a region that is dominated by background,
the $\pt(\ell\ell)$ method is also applied in a control region by restricting
\HT\ to be in the range 125--300\GeV. Here the predicted background yield is $95 \pm  16\,\text{(stat)} \pm 40 \,\text{(syst)}$ events with
$\MET> 200\GeV$, including the systematic uncertainties in the correction factors $K$ and $K_C$, and the observed yield is 59 events.

The data are displayed in the plane of \MET\ vs. \Ht\ in Fig.~\ref{fig:met_ht}.
The predicted and observed \MET\ distributions are displayed in Fig.~\ref{fig:victory}.
A summary of these results is presented in Table~\ref{tab:lightresults}.
The SF and OF observed yields in the signal regions are quoted separately, since many SUSY models
lead to enhanced production of SF lepton pairs.
For all signal regions,
the observed yield is consistent with the predictions from MC and from the background estimate
based on data. No evidence for BSM contributions to the signal regions
is observed in the light lepton channels.

\begin{table*}[htbp]
\begin{center}
\topcaption{\label{tab:lightresults}
Summary of results in the light lepton channels.
The total SM MC expected yields (MC prediction),
observed same-flavor (SF), opposite-flavor (OF), and total yields in the signal regions are indicated, as well as the
predicted yields from the \ptll\ estimate. The the expected
contributions from three benchmark SUSY scenarios are also quoted.
The first uncertainty on the \ptll\ method prediction is statistical and the second is systematic; the systematic uncertainty
is discussed in the text. The non-SM yield upper limit (UL) is a 95\% CL upper limit on the signal contribution.
}
\begin{tabular}{l|c|c|c|c}
\hline
\hline
                      &         high \MET  &        high \Ht\  & tight    &  low \Ht           \\
\hline
\hline
MC prediction         &        $30\pm1.2$    &           $31\pm0.9$ &        $12\pm0.6$     &           $4.2\pm0.3$ \\
\hline
\hline
SF yield              &                15  &               11  &                 6     &    3               \\
OF yield              &                15  &               18  &                 5     &    3               \\
\hline
\hline
\bf Total yield       &\bf             30  &\bf            29  &\bf             11     &\bf 6               \\
$\boldsymbol{\pt(\ell\ell)}$ \textbf{prediction} & $\mathbf{21\pm8.9\pm8.0}$  &  $\mathbf{22\pm7.5\pm6.9}$ & $\mathbf{11\pm5.8\pm3.8}$  & $\mathbf{12\pm4.9\pm5.7}$ \\
\hline
\hline
Observed UL       &                 26   &                  23  &                11     &               6.5     \\
Expected UL       &                 21   &                  19  &                11     &               8.6     \\

LM1                   &         $221\pm5.1$  &         $170\pm4.5$  &      $106\pm3.5$      & $6.2\pm0.9$           \\
LM3                   &          $79\pm2.4$  &          $83\pm2.5$  &      $44\pm1.8$       & $2.3\pm0.4$           \\
LM6                   &          $35\pm0.6$  &          $33\pm0.5$  &      $26\pm0.5$       & $0.6\pm0.1$           \\
LM13                  &         $133\pm5.5$  &         $113\pm5.2$  &      $65\pm3.9$       & $4.1\pm0.9$           \\
\hline
\hline
\end{tabular}
\end{center}
\end{table*}

\subsection{\texorpdfstring{Hadronic-$\Pgt$}{Hadronic-tau} channels}

In the hadronic-$\Pgt$ channels the background has two components of similar importance, events with a
genuine lepton pair
from dilepton \ttbar production and events from semi-leptonic \ttbar and \wjets\ production
with a misidentified \tauh.
Backgrounds are
estimated separately with techniques based on data control samples. Other very small contributions
from DY and diboson production with genuine lepton pairs (``MC irreducible'') are estimated from simulation.

The background with genuine lepton pairs is predicted by extending the
$\pt(\ell\ell)$ method. To translate the background prediction in the $\Pe\Pe$, $\Pe\Pgm$, and
$\Pgm\Pgm$ channels into a prediction for the $\Pe\tauh$, $\Pgm\tauh$, and $\tauh\tauh$ channels, a third correction factor is used. This correction, $K_{\Pgt} = 0.10 \pm 0.01$ for all signal regions,
is estimated from simulation and accounts for the different lepton acceptances (${\sim}0.75$),
branching fractions (${\sim}0.56$), and efficiencies (${\sim}0.24$) in hadronic-$\tau$ channels.
This procedure predicts the yield of the dilepton \ttbar\ background with genuine hadronic $\Pgt$ decays.

The background with a reconstructed $\tauh$ originating from a misidentified jet or a secondary decay is determined using a tight-to-loose ratio for $\tauh$ candidates measured in
a dijet dominated data sample, defined as $\HT > 200\GeV$ and $\MET < 20\GeV$.  Tight candidates are defined as those that
pass the full $\tauh$ selection criteria. For the definition of loose candidates, the HPS isolation criterion is replaced by a looser requirement.
The loose isolation requirement removes any \HT\ dependence of the tight-to-loose ratio; thus the measurement can
be extrapolated to the signal regions.

To determine the number of expected events including jets misidentified as $\tauh$ candidates in
the signal region,
the identification requirements for one $\tauh$ are loosened.
The obtained yields are multiplied by the probability $P_{TL}$
that a misidentified $\tauh$ candidate passes the tight $\tauh$ selection:

$$P_\mathrm{TL}(\pt, \eta) = \frac{R_\mathrm{TL}(\pt, \eta)}{1- R_\mathrm{TL}(\pt, \eta)}.$$

A summation over $P_\mathrm{TL}$ evaluated for all $\tauh$ candidates that pass the loose selection but not the tight selection gives the final background prediction in each signal region.

The method is validated in \ttbar\ simulation,
where the agreement between the predicted and true yields is within 15\%.
We correct for a 5\% bias observed in the simulation,
and assign a 15\% systematic uncertainty on
the background prediction from the tight-to-loose ratio
based on the agreement between prediction and observation
in simulation and additional control samples in data.

The results in the four signal regions are summarized in Table~\ref{tab:tauResults}.
The low-\Ht\ region includes only $\Pe\tauh$ and $\Pgm\tauh$ channels,
because the $\tauh\tauh$ trigger is inefficient in this region.
In the high-\MET\ region the $\tauh\tauh$ trigger is not fully efficient
and an efficiency correction of 3\% is applied to MC simulation.
Good agreement between predicted and observed yields is observed. No evidence
for BSM physics is observed in the hadronic-$\tau$ channels.

\begin{table*}[htbp]
\begin{center}
\topcaption{\label{tab:tauResults}
Summary of the observed and predicted yields in the four signal regions for hadronic-$\tau$ channels.
The first indicated error is statistical and the second is systematic;
the systematic uncertainties on the $R_{TL}$ ratio and \ptll\ method predictions are discussed in the text.
The non-SM yield upper limit is a 95\% CL upper limit on the signal contribution in each signal region.
}
\begin{tabular}{l|c|c|c|c}
\hline
\hline
   & high \MET & high \Ht\ & tight & low \Ht \\ \hline
$\sum\text{MC, genuine }\tauh$ & $5.8 \pm 2.3$ & $3.7 \pm 1.6$ & $2.0 \pm 1.2$ & $0.4 \pm 0.2$\\
$\sum\text{MC, misidentified }\tauh$ & $1.4 \pm 0.5$ & $2.8 \pm 1.3$ & $0.2 \pm 0.1$ & $0.2 \pm 0.1$\\
\hline
Total MC & $7.1 \pm 2.5$ & $6.5 \pm 2.3$ & $2.2 \pm 1.2$ & $0.7 \pm 0.3$\\
\hline
\hline
$\pt(\ell\ell)$ prediction & $2.1 \pm 0.9 \pm 0.8$ & $2.2 \pm 0.8 \pm 0.9$ & $1.1 \pm 0.6 \pm 0.4$ & $1.2 \pm 0.5 \pm 0.4$\\
$R_\mathrm{TL}\text{ prediction}$ & $5.1 \pm 1.7\pm 0.8$ & $3.6 \pm 1.4\pm 0.5$ & $2.7 \pm 1.3\pm 0.4$ & $< 0.9@95\% CL$\\
MC irreducible & $1.3 \pm 0.5 \pm 0.2$ & $0.7 \pm 0.3 \pm 0.1$ & $0.2 \pm 0.1 \pm 0.1$ & $0.1 \pm 0.1 \pm 0.1$\\
\hline
$\boldsymbol{\sum}$ \textbf{predictions} & $\mathbf{8.5 \pm 2.0 \pm 1.1}$ & $\mathbf{6.5 \pm 1.6 \pm 1.0}$ & $\mathbf{4.0 \pm 1.4 \pm 0.6}$ & $\mathbf{1.3 \pm 0.5 \pm 0.5}$\\
\hline\hline
\bf Total yield & \bf 8 &  \bf 5 & \bf 1 & \bf 0\\
\hline
\hline
Observed UL & $ 7.9$ &  $ 6.2$ & $ 3.7$ &  $ 3.1$\\
Expected UL & $ 8.1$ &  $ 7.2$ & $ 5.7$ &  $ 3.9$\\
$\text{LM1}$ & $32 \pm 11$ & $14 \pm 6.1$ & $8.1 \pm 4.2$ & --\\
$\text{LM3}$ & $11 \pm 4.2$ & $11 \pm 5.1$ & $8.0 \pm 4.9$ & --\\
$\text{LM6}$ & $4.5 \pm 1.5$ & $5.1 \pm 1.6$ & $4.2 \pm 1.6$ & $0.4 \pm 0.4$\\
$\text{LM13}$ & $69 \pm 17$ & $52 \pm 8.2$ & $39 \pm 9.8$ & --\\
\hline
\hline
\end{tabular}
\end{center}
\end{table*}

\begin{figure}[htbp]
\begin{center}
\includegraphics[width=\cmsFigWidth]{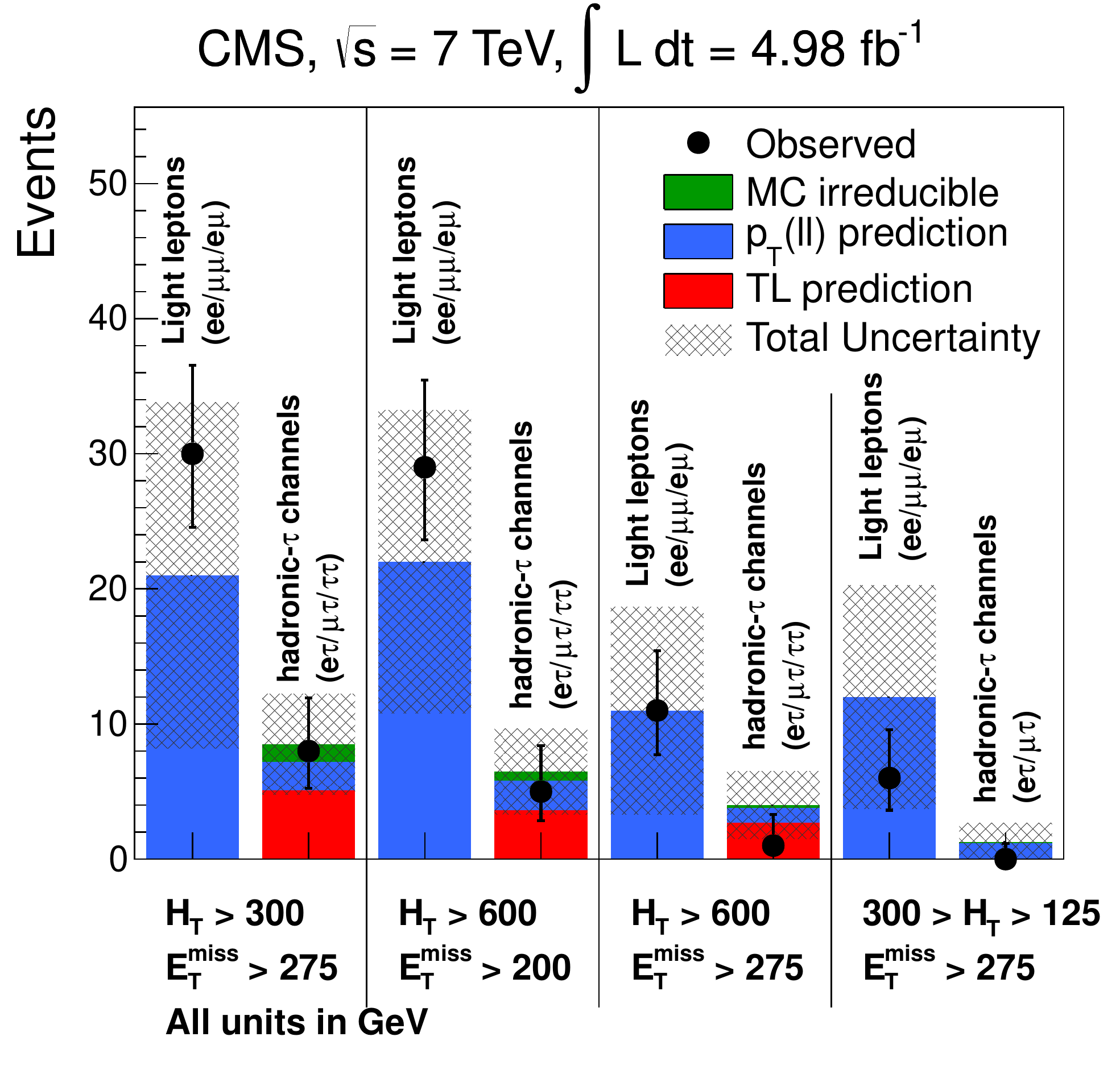}
\caption{\label{fig:tauPrediction}\protect
Summary of the background predictions from tight-to-loose ratio, $\pt(\ell\ell)$-method and MC, and observed yields in the signal regions.
}
\end{center}
\end{figure}

The results of observed yields and predicted backgrounds in all signal regions for different
lepton categories are summarized in Fig.~\ref{fig:tauPrediction}.

\section{Acceptance and Efficiency Systematic Uncertainties}
\label{sec:systematics}

The acceptance and efficiency, as well as the systematic uncertainties in these quantities,
depend on the process.
For some of the individual uncertainties, it is reasonable to quote values
based on SM control samples with kinematic properties similar to the SUSY benchmark models.
For others that depend strongly on the kinematic properties of the event, the systematic
uncertainties must be quoted model-by-model.

The systematic uncertainty in the lepton acceptance consists
of two parts: the trigger efficiency uncertainty, and the
identification and isolation uncertainty. The trigger efficiency
for two leptons of $\pt>10$\GeV, with one lepton of
$\pt>20$\GeV is measured using samples of $\Z \to \ell\ell$,
with an uncertainty of 2\%. The simulated events reproduce the lepton identification and isolation efficiencies measured in data using
samples of $\Z \to \ell\ell$ within 2\% for lepton $\pt>15$\GeV
and within 7\% (5\%) for electrons (muons) in the range $\pt=10$--15\GeV.
The uncertainty of the trigger efficiency (5\%) of the $\tauh$ triggers
is estimated with the tag-and-probe method~\cite{Htautau}.
The \tauh\ identification efficiency uncertainty is estimated to be 6\% from an independent study using a
tag-and-probe technique on $\Z \to \Pgt\Pgt$ events.
This is further validated by obtaining a
$\Z \to \Pgt\Pgt$ enhanced region showing consistency between simulation
and data.
Another significant source of systematic uncertainty is
associated with the jet and $\MET$ energy scale.  The impact
of this uncertainty is final-state dependent.  Final
states characterized by very large hadronic activity and \MET\ are
less sensitive than final states where the \MET\ and \HT\
are typically close to the minimum requirements applied to these quantities.  To be more quantitative,
we have used the method of Ref.~\cite{ref:top} to evaluate
the systematic uncertainties in the acceptance for three benchmark SUSY points.
The energies of jets in this analysis are known to within 7.5\%;
the correction accounting for the small difference between the hadronic energy scales in data and MC
is not applied~\cite{JES}.

The uncertainty on the LM1 signal efficiency in the region $\Ht\ > 300\GeV$,
\MET\ $>$ 150\GeV used to search for the kinematic edge is 6\%.
The uncertainties for the four benchmark SUSY scenarios in the signal regions used for
the counting experiments of Section~\ref{sec:datadriven} are displayed in Table~\ref{tab:jes}.
The uncertainty in the integrated luminosity is 2.2\%.%%%~\cite{ref:lumi}.

\begin{table}[htbp]
\begin{center}
\topcaption{\label{tab:jes}
Summary of the relative uncertainties in the signal efficiency due to the jet and \MET\ scale,
for the four benchmark SUSY scenarios in the signal regions used for the counting experiments of
Section~\ref{sec:datadriven}.
}
\begin{tabular}{l|cccc}
\hline
\hline
Signal Model    & high \MET  &        high \Ht  & tight    &  low \Ht           \\
\hline
LM1  &      22\%  &   33\%  &  40\% &  19\% \\
LM3  &      26\%  &   34\%  &  42\% &  18\% \\
LM6  &      11\%  &   15\%  &  19\% &  10\% \\
LM13 &      26\%  &   31\%  &  40\% &  14\% \\
\hline
\hline
\end{tabular}
\end{center}
\end{table}

\section{Limits on New Physics}
\label{sec:limit}

\subsection{Search for a kinematic edge}
\begin{figure}[tbhp]
\begin{center}
\includegraphics[width=\cmsFigWidth]{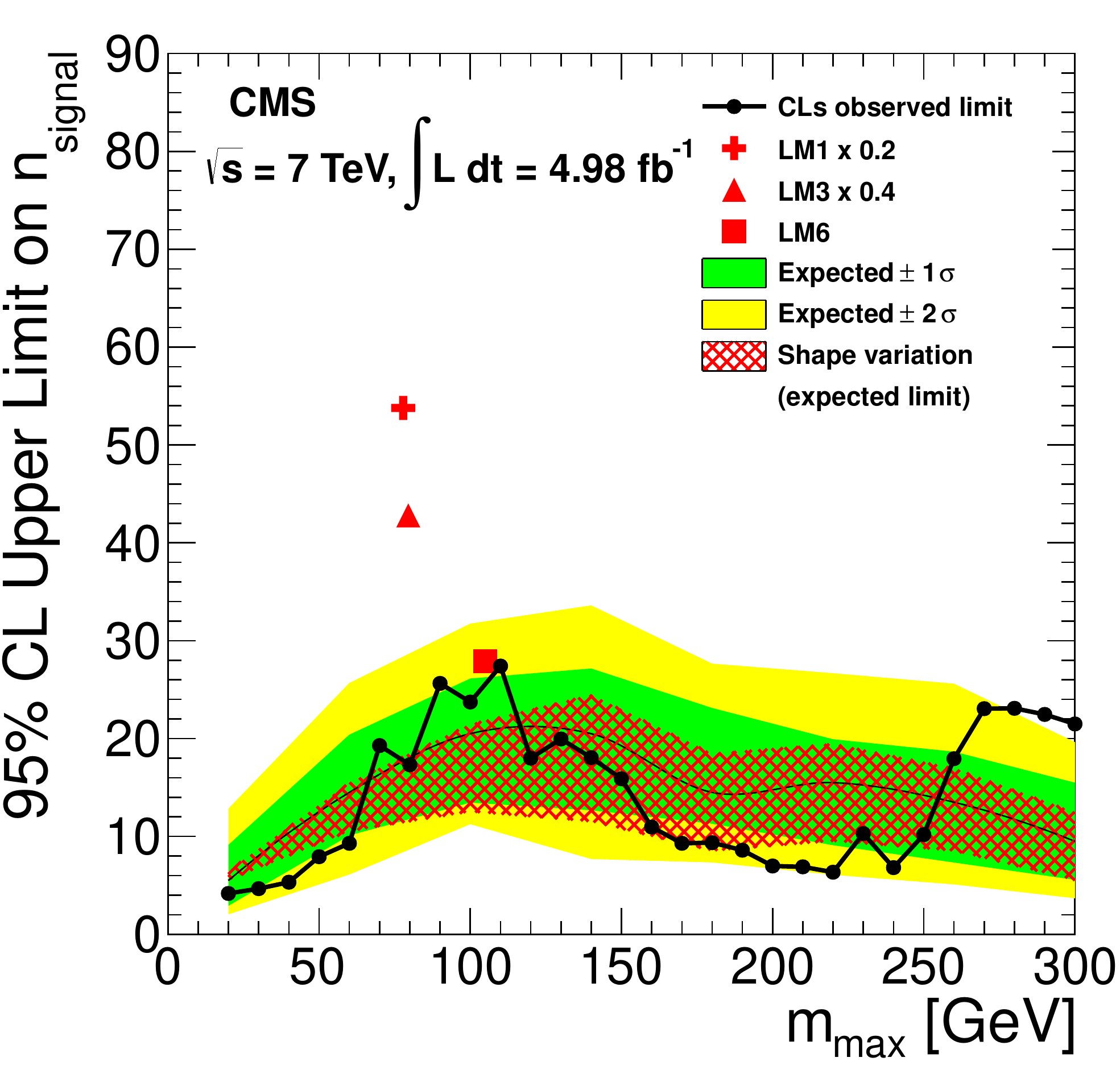}
\caption{\label{fig:limitDilmassfit}\protect
A \cls\ 95\% CL upper limit on the signal yield $n_\mathrm{S}$
as a function of the endpoint in the invariant mass spectrum $m_\text{max}$
assuming a triangular shaped signal (black dots and thick line).
The hatched band shows the variation of the expected limit (thin line)
assuming two alternate signal shapes, with the alternative expected limits corresponding
to the boundary of the hatched band.
The SUSY benchmark scenarios LM1, LM3 and LM6 are shown with their expected yields and theoretical positions of the corresponding kinematic dilepton mass edges. The LM1 (LM3) yield is scaled to 20\% (40\%) of its nominal yield.
At LM3 and LM6 a three-body decay is present; thus the shape of the kinematic edge is only approximately triangular.
}
\end{center}
\end{figure}

An upper limit on the signal yield is extracted from the fit to the dilepton mass
distribution, assuming the triangular shape ($\alpha=1$) of Eq.~(\ref{eq:triangle}).
The 95\% CL upper limit
is extracted using a hybrid frequentist-bayesian \cls\ method~\cite{ref:pdg}, including
uncertainties in the background model, resolution model and $Z$-boson yield.
We scan the position of the kinematic edge $m_\text{max}$ and extract a signal yield upper limit for each value,
as shown in Fig.~\ref{fig:limitDilmassfit}. The extracted upper limits on $n_\mathrm{S}$ vary in the range 5--30
events; these upper limits
do not depend strongly on the choice of signal shape parameter when
using two different shapes specified by a concave ($\alpha=4$) and convex curvature (hatched band).

\subsection{Search for an excess of events with large \texorpdfstring{\MET\ and \Ht}{Missing ET and HT}}

In this section we use the results of the search for events with light leptons accompanied by large \MET\ and \Ht\ reported in Section~\ref{sec:datadriven}
to exclude a region of the CMSSM parameter space. The exclusion is performed using multiple, exclusive signal regions based on the high-\MET, high-\Ht, and tight signal regions,
divided into three non-overlapping regions in the \MET\ vs. \Ht\ plane.
The results are further divided between the SF and OF final states in order to improve
the sensitivity to models with correlated dilepton production leading to an excess of SF events, yielding
a total of six signal bins, as summarized in Table~\ref{tab:lightresults_ex}.
The use of multiple, disjoint signal regions improves the sensitivity of this analysis to a specific BSM scenario.
The predicted backgrounds in the SF and OF final states are both equal to half of the total
predicted background, because the \ttbar events produce equal SF and OF yields.
The inputs to the upper limit calculation are the expected background yields and uncertainties from the \ptll\ method,
the expected signal yields and uncertainties from MC simulation, and the observed data yields in these six regions. The exclusion is performed with
the \cls\ method.
In the presence of a signal, the \ptll\ background estimate increases due to signal events populating the control
regions. To correct for this effect, for each point in the CMSSM parameter space
this expected increase is subtracted from the signal yields in our search regions.

\begin{table*}[htbp]
\begin{center}
\topcaption{\label{tab:lightresults_ex}
Summary of results in the light lepton channels used for the CMSSM exclusion of Section~\ref{sec:limit}.
Details are the same as in Table~\ref{tab:lightresults} except that
these results are divided into three non-overlapping regions defined by
$\MET > 275\GeV$, \Ht\ 300--600\GeV (SR1),
$\MET > 275\GeV$, $\Ht> 600\GeV$ (SR2, same as the ``tight'' signal region), and
\MET\ 200--275\GeV, $\Ht> 600\GeV$ (SR3).
The regions are further divided between same-flavor (SF) and opposite-flavor (OF) lepton pairs.
}
\begin{tabular}{l|ccc}
\hline
\hline
                  & SR1 & SR2 & SR3 \\
\hline
SF yield          &  9 & 6 &  5 \\
OF yield          & 10 & 5 & 13 \\
 \ptll\ prediction & $5.7\pm5.1\pm2.8$ & $5.3\pm4.1\pm1.9$ & $5.6\pm3.4\pm2.1$ \\
\hline
\hline
\end{tabular}
\end{center}
\end{table*}

\begin{figure*}[htbp]
\begin{center}
\includegraphics[width=1\linewidth]{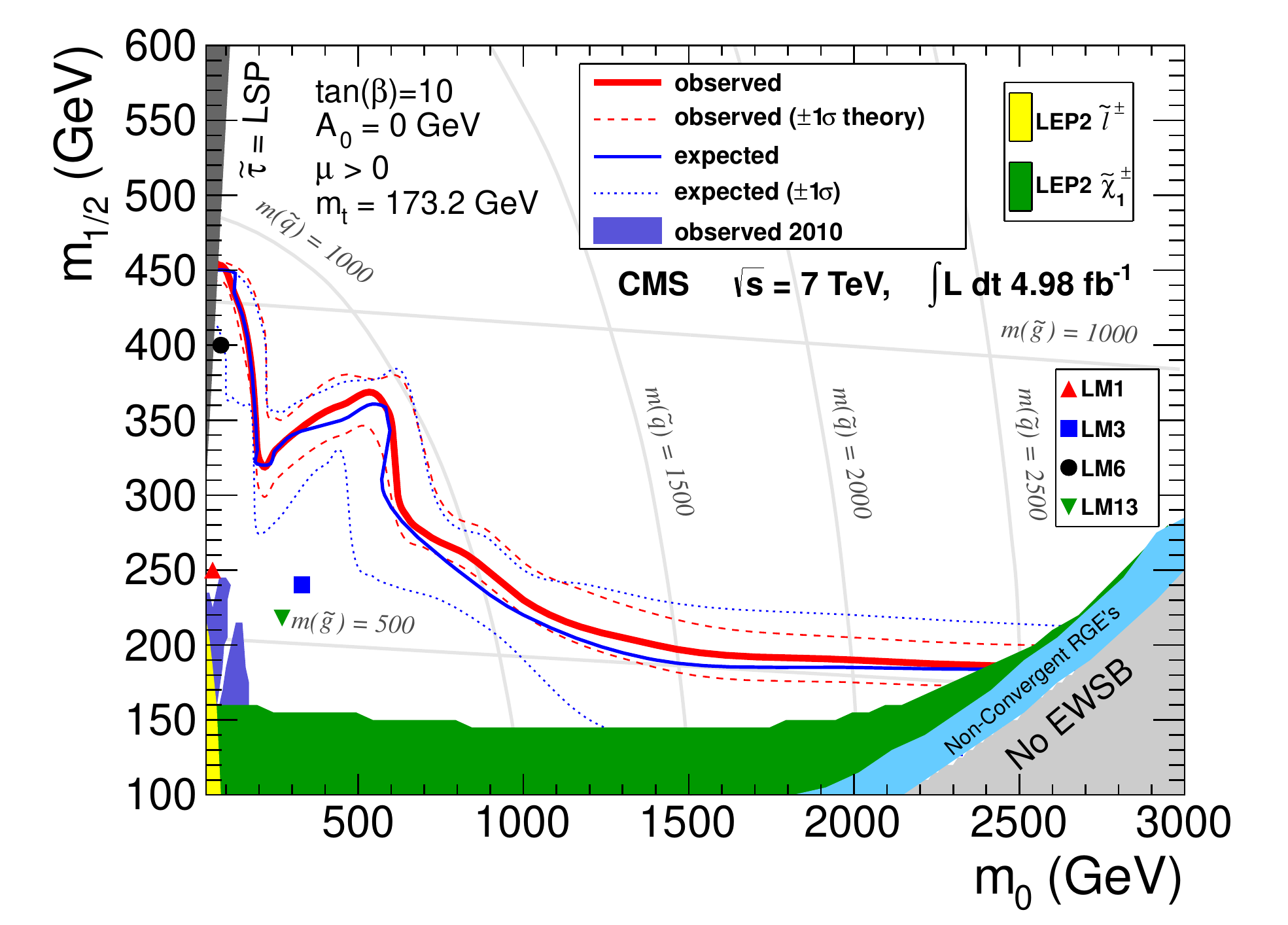}
\caption{\label{fig:cmssm}\protect
The observed 95\% CL exclusion contour (solid thick red line), the expected exclusion contour (solid thin blue line),
the variation in the observed exclusion from the variation of PDF, renormalization and factorization
scales, and $\alpha_S$ theoretical uncertainties (dashed red lines), the $\pm1\sigma$ uncertainty in the median expected exclusion
(dotted blue lines), and the observed exclusion contour based on 34 pb$^{-1}$ 2010 data in the opposite-sign dilepton channel (dark blue shaded region),
in the CMSSM $(m_0,m_{1/2})$ plane for  $\tan\beta=10$, $A_0 = 0$\GeV and $\mu > 0$.
The area below the red curve is excluded by this search. Exclusion limits obtained from
the LEP experiments are presented as shaded areas in the plot. The thin grey lines correspond to
constant squark and gluino masses. The LM benchmark SUSY scenarios are also indicated. The LM3 and LM13 benchmark scenarios
have values of  $\tan\beta$ and/or $A_0$ that differ from 10 and 0\GeV, respectively, but both are also excluded by the results of this search;
see the text of Section~\ref{sec:intro} for the full definitions of these scenarios.
}
\end{center}
\end{figure*}

The SUSY particle  spectrum is calculated using \textsc{SoftSUSY}~\cite{Allanach:2002uq}, and the
signal  events  are  generated  at  leading  order  (LO)  with  \PYTHIA6.4.22.
We use NLO  cross sections,  obtained  with the program  \PROSPINO~\cite{Beenakker:1997kx}.
Experimental uncertainties from luminosity, trigger efficiency, and lepton selection efficiency are constant across the CMSSM plane,
while the uncertainty from the hadronic energy scale is assessed separately at each CMSSM point
taking into account the bin-to-bin migration of signal events.
The variation in the observed and expected limits due to the theoretical uncertainties,
including renormalization and factorization scale, parton density functions (PDFs), and the strong coupling strength
$\alpha_S$~\cite{PDF4LHC}, are indicated in Fig.~\ref{fig:cmssm} as separate exclusion contours.
These results significantly extend the sensitivity of our previous results~\cite{Chatrchyan:2011bz}.
The  LEP-excluded regions are also indicated; these are based  on searches  for  sleptons and  charginos~\cite{LEPSUSY}.

\section{Additional Information for Model Testing}
\label{sec:outreach}

Other models of new physics in the dilepton final state can be constrained in an approximate way by simple
generator-level studies that compare the expected number of events in the data sample
corresponding to an integrated luminosity of \lumifinal\
with the upper limits from Section~\ref{sec:limit}.
The key ingredients of such studies are the kinematic requirements described
in this paper, the lepton efficiencies, and the detector responses for \HT\ and \MET.
The  trigger efficiencies for events containing $\Pe\Pe$, $\Pe\Pgm$ or $\Pgm\Pgm$ lepton pairs
are 100\%, 95\%, and 90\%, respectively. For $\Pe\tauh$, and $\Pgm\tauh$ the efficiency
is ${\sim}80\%$ \cite{Htautau}. The trigger
used for $\tauh\tauh$ final states has an efficiency of $90\%$.

\begin{figure}[htbp]
\begin{center}
\includegraphics[width=\cmsFigWidth]{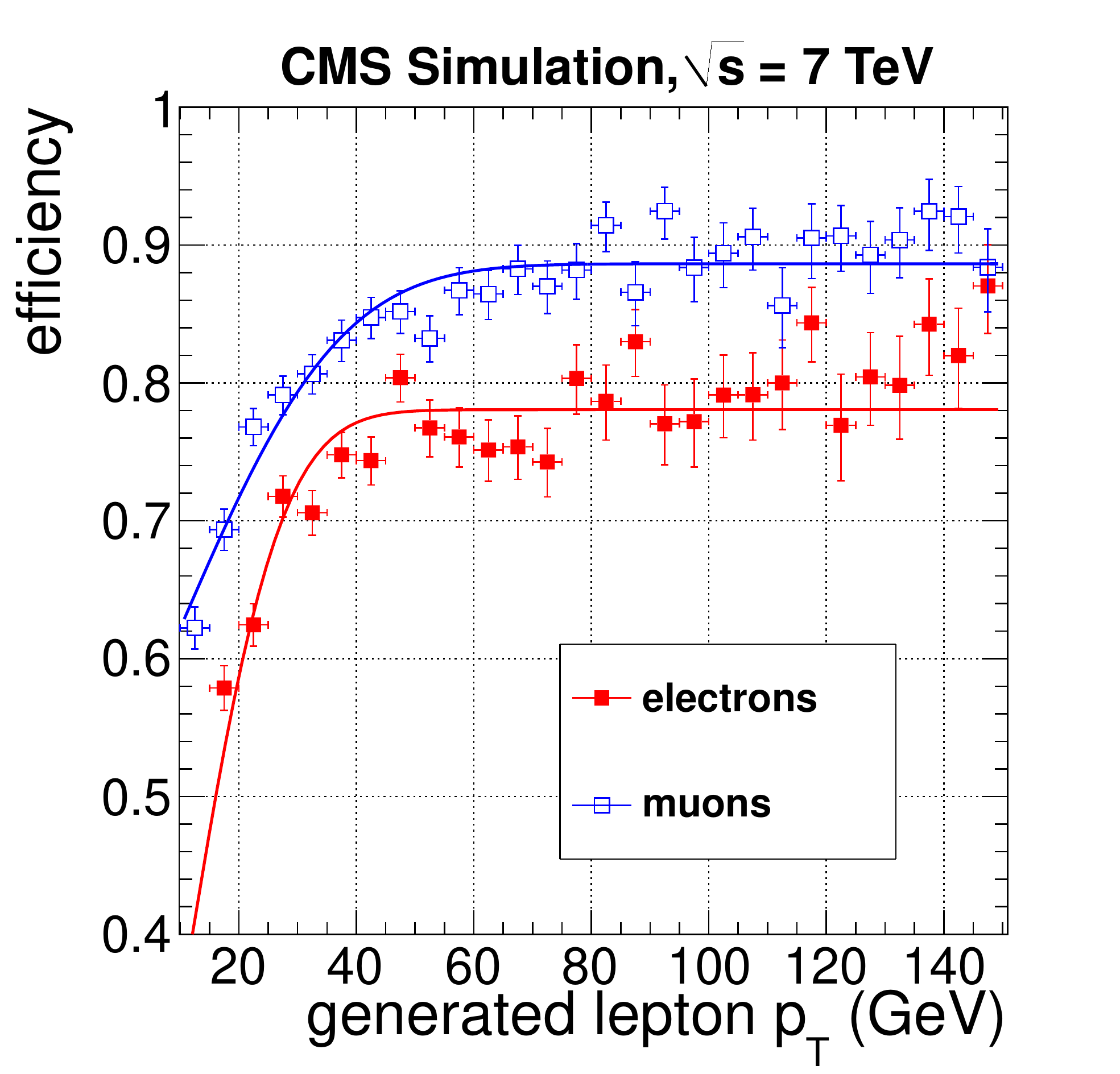}
\includegraphics[width=\cmsFigWidth]{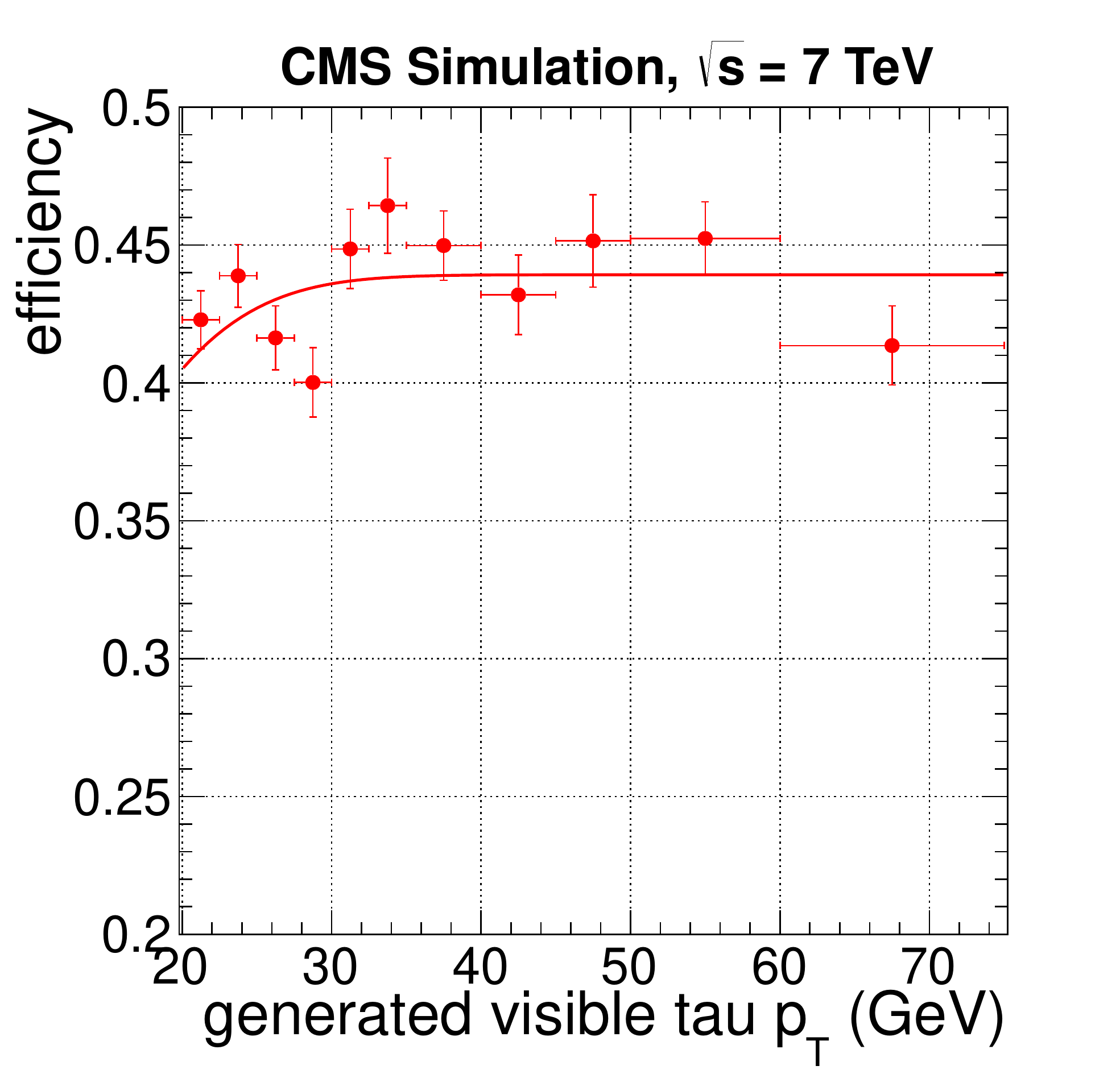}
\caption{\label{fig:lepeff}
The efficiency to pass the light lepton (\cmsLeft), and hadronic-$\Pgt$ (\cmsRight) selection as a function of the generator-level \pt (visible \tauh\ \pt).
These efficiencies are calculated using the LM6 MC benchmark.}
\end{center}
\end{figure}

We evaluate the light lepton, hadronic-$\Pgt$, \MET, and \Ht\ selection efficiencies using the LM6 benchmark model, but these
efficiencies do not depend strongly on the choice of model.
Jets at the generator-level are approximated as quarks or gluons produced prior to the parton showering step satisfying
$\pt> 30\GeV$ and $|\eta|<3$.
Generator-level leptons are required to satisfy $\pt> 10\GeV$ and $|\eta|<2.5$ and not
to overlap with a generator-level jet within $\Delta R<0.4$. For generator level
\tauh\ the visible decay products are required to satisfy the tighter \pt\ $>$ 20
GeV and $|\eta|<2.1$ selection.
The generator-level \MET\ is the absolute value of the vector sum of the transverse momenta of invisible
particles, e.g., neutrinos and lightest supersymmetric particles. The lepton selection efficiencies as a function of generator-level \pt\ are displayed
in Fig.~\ref{fig:lepeff}. The efficiency dependence can be parameterized as a function of \pt\ as

\begin{equation}
\label{eq:erf1}
f(p_T) = \epsilon_{\infty} \{\erf[(\pt - C)/\sigma] \} + \epsilon_C \{ 1 - \erf[(\pt - C)/\sigma]\},
\end{equation}

where $\erf$ indicates the error function, $\epsilon_{\infty}$ gives the value of the efficiency plateau at high momenta,
$C$ is equal to 10\GeV, $\epsilon_C$ gives the value of the efficiency at $\pt=C$,
and $\sigma$ describes how fast the transition is.
The parameterization is summarized in Table~\ref{tab:lepeffLM6fit} for electrons, muons, and taus.

\begin{table}[htbp]
\begin{center}
\topcaption{\label{tab:lepeffLM6fit}Values of the fitted parameters in Eq.~(\ref{eq:erf1}) for the lepton selection efficiencies of Fig.~\ref{fig:lepeff}.}
\begin{tabular}{l|ccc}\hline\hline
Parameter		& $\Pe$	        	& $\Pgm$         & \tauh\	\\ \hline
$C$      		& 10\GeV	        & 10\GeV	& 10\GeV	\\
$\epsilon_{\infty}$	& $0.78$		& $0.89$        & $0.44$        \\
$\epsilon_{C}$		& $0.34$		& $0.62$        & $0.31$        \\
$\sigma$         	& $18$\GeV		& $30$\GeV	& $13$\GeV	\\
\hline\hline
\end{tabular}
\end{center}
\end{table}

\begin{figure}[htbp]
\begin{center}
\includegraphics[width=\cmsFigWidth]{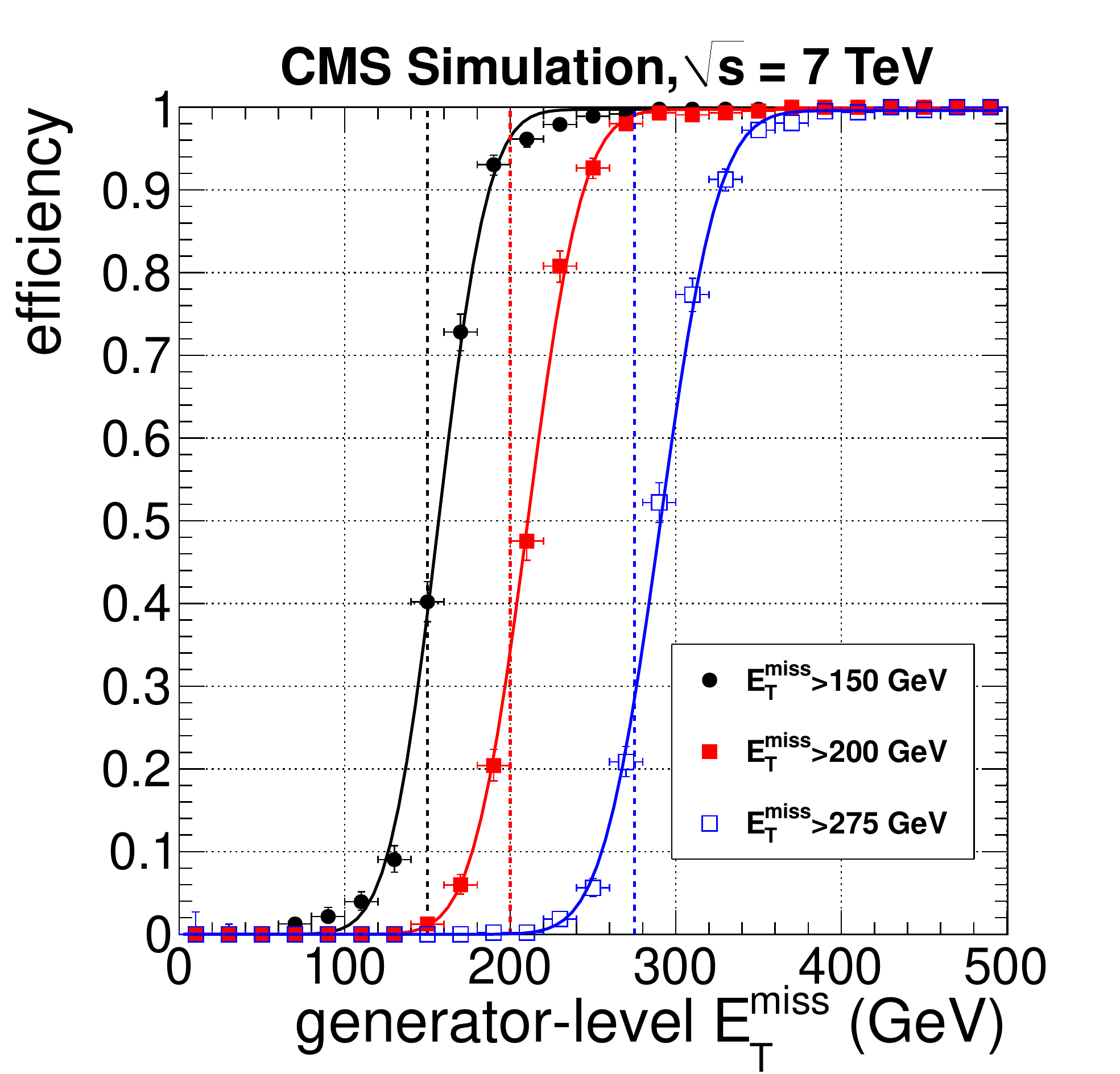}
\includegraphics[width=\cmsFigWidth]{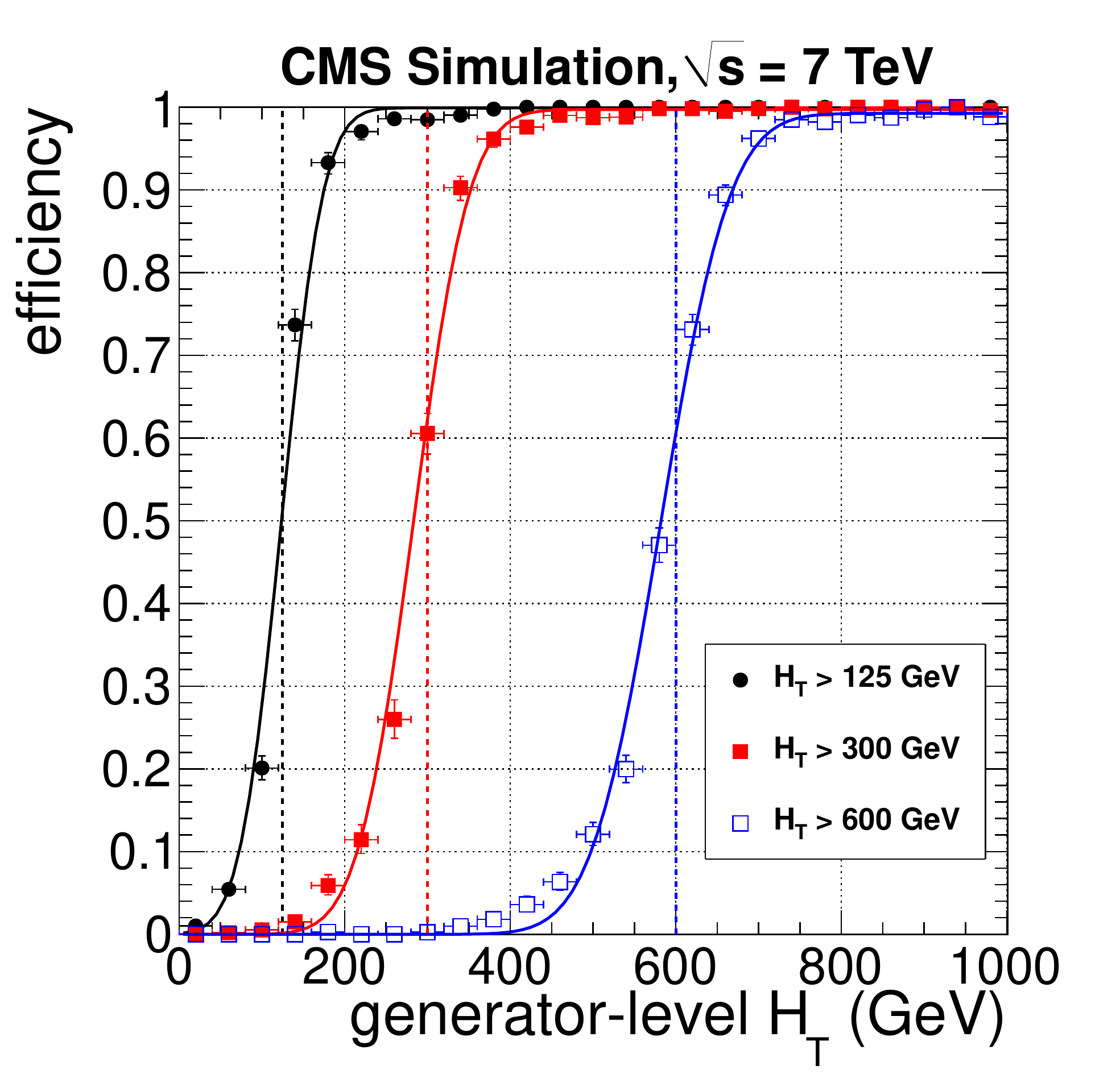}
\caption{\label{fig:response}
The efficiency to pass the signal region \MET\ (\cmsLeft), and \Ht\ (\cmsRight) requirements as a function of the generator-level quantities.
The vertical lines represent the requirements applied to the reconstruction-level quantities. These efficiencies are calculated
using the LM6 MC benchmark, but they do not depend strongly on the underlying physics.
}
\end{center}
\end{figure}

The \MET\ and \Ht\ selection efficiencies are displayed in Fig.~\ref{fig:response} as a function of the generator-level quantities.
These efficiencies are parameterized using the function:

\begin{equation}
\label{eq:erf}
f(x) = \frac{\epsilon_{\infty}}{2}~( \erf((x-C)/\sigma) + 1 ),
\end{equation}

where $\epsilon_{\infty}$ gives the value of the efficiency plateau at high $x$,
$C$ is the value of $x$ at which the efficiency is equal to 50\%,
and $\sigma$ describes how fast the transition is.
The values of the fitted parameters are quoted in Table~\ref{tab:fit}.

\begin{table*}[htbp]
\begin{center}
\caption{\label{tab:fit}
Values of the fitted parameters in Eq.~(\ref{eq:erf}) for the \MET\ and \Ht\ selection efficiencies of Fig.~\ref{fig:response}.
}
\begin{tabular}{lccccc}
\hline
\hline
Parameter               &  $\MET> 150\GeV$   &   $\MET> 200\GeV$   &  $\MET>275\GeV$   &    \\
\hline
$\epsilon_{\infty}$       &     1.00             &    1.00               &       1.00           &    \\
$C$                     &   157\GeV            &    211\GeV            &       291\GeV        &    \\
$\sigma$                &   33\GeV             &    37\GeV             &        39\GeV        &    \\
\hline
\hline
Parameter               &   $\Ht> 125\GeV$    &  $\Ht>300\GeV$    &   $\Ht> 600\GeV $  &    \\
\hline
$\epsilon_{\infty}$      &    1.00               &       1.00           &  0.99     \\
$C$                    &    124\GeV            &       283\GeV        &  582\GeV  \\
$\sigma$               &    ~56\GeV            &       ~75\GeV        &  ~93\GeV  \\
\hline
\hline
\end{tabular}
\end{center}
\end{table*}

This efficiency model has been validated by comparing the yields from the full reconstruction with the expected
yields using generator-level information only and the efficiencies quoted above.  %, for LM1, LM3, and LM6.
In addition to the LM1, LM3, LM6 and LM13 benchmarks considered throughout this paper, we have tested several
additional benchmarks (LM2, LM4, LM5, LM7, and LM8)~\cite{PTDR2}.
In general we observe agreement between full reconstruction and the efficiency model within approximately 15\%.
\section{Summary}
\label{sec:conclusion}
We have presented a search for physics beyond the standard model in the opposite-sign dilepton final state using
a data sample of proton-proton collisions at a center-of-mass energy of 7\TeV. The data sample corresponds to an integrated
luminosity of \lumifinal, and was collected with the CMS detector in 2011.
Two complementary search strategies have been performed. The first focuses on models with a specific dilepton
production mechanism leading to a characteristic kinematic edge in the dilepton mass distribution, and
the second focuses on dilepton events accompanied by large missing transverse energy and significant hadronic activity.
This work is motivated by many models of BSM physics, such as supersymmetric models or models with universal extra dimensions.
In the absence of evidence for BSM physics, we set upper limits on the BSM contributions to yields in the signal regions.
Additional information has been provided to allow testing whether specific models of new physics are excluded
by these results.
The presented result is the most stringent limit to date from the opposite-sign dilepton final state
accompanied by large missing transverse energy and hadronic activity.

\section*{Acknowledgements}\label{sec:ack}
We congratulate our colleagues in the CERN accelerator departments for the excellent performance of the LHC machine. We thank the technical and administrative staff at CERN and other CMS institutes, and acknowledge support from: FMSR (Austria); FNRS and FWO (Belgium); CNPq, CAPES, FAPERJ, and FAPESP (Brazil); MES (Bulgaria); CERN; CAS, MoST, and NSFC (China); COLCIENCIAS (Colombia); MSES (Croatia); RPF (Cyprus); MoER, SF0690030s09 and ERDF (Estonia); Academy of Finland, MEC, and HIP (Finland); CEA and CNRS/IN2P3 (France); BMBF, DFG, and HGF (Germany); GSRT (Greece); OTKA and NKTH (Hungary); DAE and DST (India); IPM (Iran); SFI (Ireland); INFN (Italy); NRF and WCU (Korea); LAS (Lithuania); CINVESTAV, CONACYT, SEP, and UASLP-FAI (Mexico); MSI (New Zealand); PAEC (Pakistan); MSHE and NSC (Poland); FCT (Portugal); JINR (Armenia, Belarus, Georgia, Ukraine, Uzbekistan); MON, RosAtom, RAS and RFBR (Russia); MSTD (Serbia); MICINN and CPAN (Spain); Swiss Funding Agencies (Switzerland); NSC (Taipei); TUBITAK and TAEK (Turkey); STFC (United Kingdom); DOE and NSF (USA). Individuals have received support from the Marie-Curie programme and the European Research Council (European Union); the Leventis Foundation; the A. P. Sloan Foundation; the Alexander von Humboldt Foundation; the Belgian Federal Science Policy Office; the Fonds pour la Formation \`a la Recherche dans l'Industrie et dans l'Agriculture (FRIA-Belgium); the Agentschap voor Innovatie door Wetenschap en Technologie (IWT-Belgium); the Council of Science and Industrial Research, India; and the HOMING PLUS programme of Foundation for Polish Science, cofinanced from European Union, Regional Development Fund.

\vspace*{12pt}
\bibliography{auto_generated}   % will be created by the tdr script.

\providecommand{\href}[2]{#2}\begingroup\raggedright\begin{thebibliography}{10}%
\makeatletter
\providecommand{\hrefCMSnoop }[0]{\@secondoftwo}%
\makeatother
\providecommand{\doi}{\texttt{doi:}\begingroup \urlstyle{tt}\Url}

\bibitem{CMS:2008zzk}
\hrefCMSnoop {} {{ CMS} Collaboration, ``{The CMS experiment at the CERN
  LHC}'',} \textit{ JINST} \textbf{ 3} (2008) S08004,
\href{http://dx.doi.org/10.1088/1748-0221/3/08/S08004}{\doi{10.1088/1748-0221/3/08/S08004}}.
%%CITATION = JINST,3,S08004;%%.

\bibitem{Chatrchyan:2011bz}
\hrefCMSnoop {} {{ CMS} Collaboration, ``{Search for physics beyond the
  standard model in opposite-sign dilepton events at $\sqrt{s} = 7$ TeV}'',}
  \textit{ JHEP} \textbf{ 06} (2011) 026,
  \href{http://dx.doi.org/10.1007/JHEP06(2011)026}{\doi{10.1007/JHEP06(2011)026}},
\href{http://www.arXiv.org/abs/1103.1348}{\texttt{ arXiv:1103.1348}}.
%%CITATION = ARXIV:1103.1348;%%.

\bibitem{DM1}
J.~Ellis\hrefCMSnoop {} { {et~al.}, ``{Exploration of the MSSM with
  nonuniversal Higgs masses}'',} \textit{ Nucl. Phys. B} \textbf{ 652} (2003)
  259,
  \href{http://dx.doi.org/10.1016/S0550-3213(02)01144-6}{\doi{10.1016/S0550-3213(02)01144-6}},
\href{http://www.arXiv.org/abs/hep-ph/0210205}{\texttt{ arXiv:hep-ph/0210205}}.
%%CITATION = HEP-PH/0210205;%%.

\bibitem{DM2}
J.~Ellis\hrefCMSnoop {} { {et~al.}, ``{Supersymmetric dark matter in light of
  WMAP}'',} \textit{ Phys. Lett. B} \textbf{ 565} (2003) 176,
  \href{http://dx.doi.org/10.1016/S0370-2693(03)00765-2}{\doi{10.1016/S0370-2693(03)00765-2}},
\href{http://www.arXiv.org/abs/hep-ph/0303043}{\texttt{ arXiv:hep-ph/0303043}}.
%%CITATION = HEP-PH/0303043;%%.

\bibitem{DM3}
D.~Auto\hrefCMSnoop {} { {et~al.}, ``{Y}ukawa coupling unification in
  supersymmetric models'',} \textit{ JHEP} \textbf{ 06} (2003) 023,
  \href{http://dx.doi.org/10.1088/1126-6708/2003/06/023}{\doi{10.1088/1126-6708/2003/06/023}},
\href{http://www.arXiv.org/abs/hep-ph/0302155}{\texttt{ arXiv:hep-ph/0302155}}.
%%CITATION = HEP-PH/0302155;%%.

\bibitem{DM4}
A.~Bottin\hrefCMSnoop {} { {et~al.}, ``{Lower bound on the neutralino mass from
  new data on CMB and implications for relic neutralinos}'',} \textit{ Phys.
  Rev. D} \textbf{ 68} (2003) 043506,
  \href{http://dx.doi.org/10.1103/PhysRevD.68.043506}{\doi{10.1103/PhysRevD.68.043506}},
\href{http://www.arXiv.org/abs/hep-ph/0304080}{\texttt{ arXiv:hep-ph/0304080}}.
%%CITATION = HEP-PH/0304080;%%.

\bibitem{Martin:1997ns}
\hrefCMSnoop {} {S.~P. Martin, ``{A Supersymmetry primer}'',}.
  \href{http://www.arXiv.org/abs/hep-ph/9709356}{\texttt{
  arXiv:hep-ph/9709356}}.

\bibitem{Wess:1974tw}
\hrefCMSnoop {} {J.~Wess and B.~Zumino, ``{Supergauge Transformations in
  Four-Dimensions}'',} \textit{ Nucl. Phys. B} \textbf{ 70} (1974) 39,
  \href{http://dx.doi.org/10.1016/0550-3213(74)90355-1}{\doi{10.1016/0550-3213(74)90355-1}}.

\bibitem{UEDColl}
M.~Battaglia\hrefCMSnoop {} { {et~al.}, ``Contrasting supersymmetry and
  universal extra dimensions at colliders'',} (2005).
\href{http://www.arXiv.org/abs/hep-ph/0507284}{\texttt{ arXiv:hep-ph/0507284}}.
%%CITATION = HEP-PH/0507284;%%.

\bibitem{ref:RA1}
\hrefCMSnoop {} {V.~Khachatryan {et~al.}, ``{Search for supersymmetry in pp
  collisions at 7 TeV in events with jets and missing transverse energy}'',}
  \textit{ Phys. Lett. B} \textbf{ 698} (2011) 196,
  \href{http://dx.doi.org/10.1016/j.physletb.2011.03.021}{\doi{10.1016/j.physletb.2011.03.021}},
  \href{http://www.arXiv.org/abs/1101.1628}{\texttt{ arXiv:1101.1628}}.

\bibitem{ref:SS}
\hrefCMSnoop {} {{ CMS} Collaboration, ``{Search for new physics with same-sign
  isolated dilepton events with jets and missing transverse energy at the
  LHC}'',} \textit{ JHEP} \textbf{ 06} (2011) 077,
  \href{http://dx.doi.org/10.1007/JHEP06(2011)077}{\doi{10.1007/JHEP06(2011)077}},
\href{http://www.arXiv.org/abs/1104.3168}{\texttt{ arXiv:1104.3168}}.
%%CITATION = ARXIV:1104.3168;%%.

\bibitem{Aad:2011cwa}
\hrefCMSnoop {} {{ ATLAS} Collaboration, ``{Searches for supersymmetry with the
  ATLAS detector using final states with two leptons and missing transverse
  momentum in $\sqrt{s} = 7$ TeV proton-proton collisions}'',} \textit{ Phys.
  Lett. B} \textbf{ 709} (2012) 137,
  \href{http://dx.doi.org/10.1016/j.physletb.2012.01.076}{\doi{10.1016/j.physletb.2012.01.076}},
\href{http://www.arXiv.org/abs/1110.6189}{\texttt{ arXiv:1110.6189}}.
%%CITATION = ARXIV:1110.6189;%%.

\bibitem{ATLAS:2012ag}
\hrefCMSnoop {} {{ ATLAS} Collaboration, ``{Search for events with large
  missing transverse momentum, jets, and at least two tau leptons in 7 TeV
  proton-proton collision data with the ATLAS detector}'',} \textit{ Phys.
  Lett. B} \textbf{ 714} (2012) 180,
  \href{http://dx.doi.org/10.1016/j.physletb.2012.06.055}{\doi{10.1016/j.physletb.2012.06.055}},
\href{http://www.arXiv.org/abs/1203.6580}{\texttt{ arXiv:1203.6580}}.
%%CITATION = ARXIV:1203.6580;%%.

\bibitem{ref:top}
\hrefCMSnoop {} {{ CMS} Collaboration, ``{First measurement of the cross
  section for top-quark pair production in proton-proton collisions at
  $\sqrt(s)=7$~TeV}'',} \textit{ Phys. Lett. B} \textbf{ 695} (2011) 424,
  \href{http://dx.doi.org/10.1016/j.physletb.2010.11.058}{\doi{10.1016/j.physletb.2010.11.058}},
  \href{http://www.arXiv.org/abs/1010.5994}{\texttt{ arXiv:1010.5994}}.

\bibitem{edge}
I.~Hinchliffe\hrefCMSnoop {} { {et~al.}, ``{Precision SUSY measurements at CERN
  LHC}'',} \textit{ Phys. Rev. D} \textbf{ 55} (1997) 5520,
  \href{http://dx.doi.org/10.1103/PhysRevD.55.5520}{\doi{10.1103/PhysRevD.55.5520}}.

\bibitem{CMSSM}
G.~L. Kane\hrefCMSnoop {} { {et~al.}, ``Study of constrained minimal
  supersymmetry'',} \textit{ Phys. Rev. D} \textbf{ 49} (1994) 6173,
  \href{http://dx.doi.org/10.1103/PhysRevD.49.6173}{\doi{10.1103/PhysRevD.49.6173}},
  \href{http://www.arXiv.org/abs/hep-ph/9312272}{\texttt{
  arXiv:hep-ph/9312272}}.

\bibitem{CMSSM2}
\hrefCMSnoop {} {A.~H. Chamseddine, R.~L. Arnowitt, and P.~Nath, ``{Locally
  Supersymmetric Grand Unification}'',} \textit{ Phys. Rev. Lett.} \textbf{ 49}
  (1982) 970,
  \href{http://dx.doi.org/10.1103/PhysRevLett.49.970}{\doi{10.1103/PhysRevLett.49.970}}.

\bibitem{PTDR2}
\hrefCMSnoop {} {{CMS Collaboration}, ``{CMS} technical design report, volume
  {II}: {Physics} performance'',} \textit{ J. Phys. G} \textbf{ 34} (2007) 995,
\href{http://dx.doi.org/10.1088/0954-3899/34/6/S01}{\doi{10.1088/0954-3899/34/6/S01}}.
%%CITATION = JPHGB,G34,995;%%.

\bibitem{CMS-PAS-PFT-10-002}
\href {http://cdsweb.cern.ch/record/1279341} {{ CMS} Collaboration,
  ``Commissioning of the Particle-Flow Reconstruction in Minimum-Bias and Jet
  Events from {\Pp\Pp} Collisions at 7 {TeV}'',} CMS Physics Analysis Summary
  CMS-PAS-PFT-10-002, (2010).

\bibitem{Pythia}
\hrefCMSnoop {} {T.~Sj\"ostrand, S.~Mrenna, and P.~Z. Skands, ``{PYTHIA 6.4
  physics and manual}'',} \textit{ JHEP} \textbf{ 05} (2006) 026,
  \href{http://dx.doi.org/10.1088/1126-6708/2006/05/026}{\doi{10.1088/1126-6708/2006/05/026}},
  \href{http://www.arXiv.org/abs/hep-ph/0603175}{\texttt{
  arXiv:hep-ph/0603175}}.

\bibitem{Madgraph}
\hrefCMSnoop {} {J.~Alwall, ``{MadGraph/MadEvent v4: the new web
  generation}'',} \textit{ JHEP} \textbf{ 09} (2007) 028,
  \href{http://dx.doi.org/10.1088/1126-6708/2007/09/028}{\doi{10.1088/1126-6708/2007/09/028}}.

\bibitem{POWHEG}
\hrefCMSnoop {} {S.~Frixione, P.~Nason, and C.~Oleari, ``{Matching NLO QCD
  computations with parton shower simulations: the POWHEG method}'',} \textit{
  JHEP} \textbf{ 11} (2007) 070,
  \href{http://dx.doi.org/10.1088/1126-6708/2007/11/070}{\doi{10.1088/1126-6708/2007/11/070}},
\href{http://www.arXiv.org/abs/0709.2092}{\texttt{ arXiv:0709.2092}}.
%%CITATION = ARXIV:0709.2092;%%.

\bibitem{cteq66}
P.~M. Nadolsky\hrefCMSnoop {} { {et~al.}, ``{Implications of CTEQ global
  analysis for collider observables}'',} \textit{ Phys. Rev. D} \textbf{ 78}
  (2008) 013004,
  \href{http://dx.doi.org/10.1103/PhysRevD.78.013004}{\doi{10.1103/PhysRevD.78.013004}},
\href{http://www.arXiv.org/abs/0802.0007}{\texttt{ arXiv:0802.0007}}.
%%CITATION = 0802.0007;%%.

\bibitem{Chatrchyan:2011id}
\hrefCMSnoop {} {{ CMS} Collaboration, ``{Measurement of the Underlying Event
  Activity at the LHC with $\sqrt{s}= 7$ TeV and Comparison with $\sqrt{s} =
  0.9$ TeV}'',} \textit{ JHEP} \textbf{ 1109} (2011) 109,
  \href{http://dx.doi.org/10.1007/JHEP09(2011)109}{\doi{10.1007/JHEP09(2011)109}},
\href{http://www.arXiv.org/abs/1107.0330}{\texttt{ arXiv:1107.0330}}.
%%CITATION = ARXIV:1107.0330;%%.

\bibitem{Geant}
\hrefCMSnoop {} {{ GEANT4} Collaboration, ``{GEANT4}--a simulation toolkit'',}
  \textit{ Nucl. Instrum. Meth. A} \textbf{ 506} (2003) 250,
  \href{http://dx.doi.org/10.1016/S0168-9002(03)01368-8}{\doi{10.1016/S0168-9002(03)01368-8}}.

\bibitem{Khachatryan:2010pw}
\hrefCMSnoop {} {{ CMS} Collaboration, ``{CMS Tracking Performance Results from
  early LHC Operation}'',} \textit{ Eur. Phys. J. C} \textbf{ 70} (2010) 1165,
  \href{http://dx.doi.org/10.1140/epjc/s10052-010-1491-3}{\doi{10.1140/epjc/s10052-010-1491-3}},
\href{http://www.arXiv.org/abs/1007.1988}{\texttt{ arXiv:1007.1988}}.
%%CITATION = ARXIV:1007.1988;%%.

\bibitem{TauPAS}
\href {http://cdsweb.cern.ch/record/1337004} {{ CMS} Collaboration,
  ``Performance of tau reconstruction algorithms in 2010 data collected with
  {CMS}'',} CMS Physics Analysis Summary CMS-PAS-TAU-11-001, (2011).

\bibitem{antikt}
\hrefCMSnoop {} {M.~Cacciari, G.~P. Salam, and G.~Soyez, ``The anti-$k_t$ jet
  clustering algorithm'',} \textit{ JHEP} \textbf{ 04} (2008) 063,
  \href{http://dx.doi.org/10.1088/1126-6708/2008/04/063}{\doi{10.1088/1126-6708/2008/04/063}},
  \href{http://www.arXiv.org/abs/0802.1189}{\texttt{ arXiv:0802.1189}}.

\bibitem{top1}
\hrefCMSnoop {} {{ CMS} Collaboration, ``{First Measurement of the Cross
  Section for Top-Quark Pair Production in Proton-Proton Collisions at
  $\sqrt{s}=7$ TeV}'',} \textit{ Phys. Lett. B} \textbf{ 695} (2011) 424,
  \href{http://dx.doi.org/10.1016/j.physletb.2010.11.058}{\doi{10.1016/j.physletb.2010.11.058}},
\href{http://www.arXiv.org/abs/1010.5994}{\texttt{ arXiv:1010.5994}}.
%%CITATION = ARXIV:1010.5994;%%.

\bibitem{top2}
\hrefCMSnoop {} {{ CMS} Collaboration, ``{Measurement of the \ttbar production
  cross section and the top quark mass in the dilepton channel in pp collisions
  at $\sqrt{s} =7$ TeV}'',} \textit{ JHEP} \textbf{ 07} (2011) 049,
  \href{http://dx.doi.org/10.1007/JHEP07(2011)049}{\doi{10.1007/JHEP07(2011)049}},
\href{http://www.arXiv.org/abs/1105.5661}{\texttt{ arXiv:1105.5661}}.
%%CITATION = ARXIV:1105.5661;%%.

\bibitem{top3}
\hrefCMSnoop {} {{ CMS} Collaboration, ``{Measurement of the \ttbar production
  cross section in $pp$ collisions at 7 Tev in lepton+jets events using
  $b$-quark jet identification}'',} \textit{ Phys. Rev. D} \textbf{ 84} (2011)
  092004,
  \href{http://dx.doi.org/10.1103/PhysRevD.84.092004}{\doi{10.1103/PhysRevD.84.092004}},
\href{http://www.arXiv.org/abs/1108.3773}{\texttt{ arXiv:1108.3773}}.
%%CITATION = ARXIV:1108.3773;%%.

\bibitem{LEE}
\hrefCMSnoop {} {E.~Gross and O.~Vitells, ``{Trial factors for the look
  elsewhere effect in high energy physics}'',} \textit{ Eur. Phys. J. C}
  \textbf{ 70} (2010) 525,
  \href{http://dx.doi.org/10.1140/epjc/s10052-010-1470-8}{\doi{10.1140/epjc/s10052-010-1470-8}},
\href{http://www.arXiv.org/abs/1005.1891}{\texttt{ arXiv:1005.1891}}.
%%CITATION = 1005.1891;%%.

\bibitem{ref:victory}
\hrefCMSnoop {} {V.~Pavlunin, ``{Modeling missing transverse energy in V+jets
  at CERN LHC}'',} \textit{ Phys. Rev. D} \textbf{ 81} (2010) 035005,
  \href{http://dx.doi.org/10.1103/PhysRevD.81.035005}{\doi{10.1103/PhysRevD.81.035005}},
  \href{http://www.arXiv.org/abs/0906.5016}{\texttt{ arXiv:0906.5016}}.

\bibitem{Wpolarization}
J.~A. Aguilar-Saavedra\hrefCMSnoop {} { {et~al.}, ``{Probing anomalous Wtb
  couplings in top pair decays}'',} \textit{ Eur. Phys. J. C} \textbf{ 50}
  (2007) 519,
  \href{http://dx.doi.org/10.1140/epjc/s10052-007-0289-4}{\doi{10.1140/epjc/s10052-007-0289-4}},
  \href{http://www.arXiv.org/abs/hep-ph/0605190}{\texttt{
  arXiv:hep-ph/0605190}}.

\bibitem{Wpolarization2}
\hrefCMSnoop {} {A.~Czarnecki, J.~G. Korner, and J.~H. Piclum, ``{Helicity
  fractions of W bosons from top quark decays at NNLO in QCD}'',} \textit{
  Phys. Rev. D} \textbf{ 81} (2010) 111503,
  \href{http://dx.doi.org/10.1103/PhysRevD.81.111503}{\doi{10.1103/PhysRevD.81.111503}},
  \href{http://www.arXiv.org/abs/1005.2625}{\texttt{ arXiv:1005.2625}}.

\bibitem{Wpolarization3}
\hrefCMSnoop {} {{ CDF} Collaboration, ``{Measurement of W-Boson Polarization
  in Top-quark Decay in $p\bar{p}$ Collisions at $\sqrt{s} = 1.96$ TeV}'',}
  \textit{ Phys. Rev. Lett.} \textbf{ 105} (2010) 042002,
  \href{http://dx.doi.org/10.1103/PhysRevLett.105.042002}{\doi{10.1103/PhysRevLett.105.042002}},
\href{http://www.arXiv.org/abs/1003.0224}{\texttt{ arXiv:1003.0224}}.
%%CITATION = ARXIV:1003.0224;%%.

\bibitem{Htautau}
\hrefCMSnoop {} {{CMS Collaboration}, ``Search for neutral {H}iggs bosons
  decaying to tau pairs in pp collisions at $\sqrt{s}$ = 7 {TeV}'',} \textit{
  Phys. Lett. B} (2012) 68,
  \href{http://dx.doi.org/10.1016/j.physletb.2012.05.028}{\doi{10.1016/j.physletb.2012.05.028}},
  \href{http://www.arXiv.org/abs/1202.4083}{\texttt{ arXiv:1202.4083}}.

\bibitem{JES}
\hrefCMSnoop {} {{ CMS} Collaboration, ``{Determination of Jet Energy
  Calibration and Transverse Momentum Resolution in CMS}'',} \textit{ JINST}
  \textbf{ 6} (2011) P11002,
  \href{http://dx.doi.org/10.1088/1748-0221/6/11/P11002}{\doi{10.1088/1748-0221/6/11/P11002}},
\href{http://www.arXiv.org/abs/1107.4277}{\texttt{ arXiv:1107.4277}}.
%%CITATION = ARXIV:1107.4277;%%.

\bibitem{ref:pdg}
\hrefCMSnoop {} {{ Particle Data Group} Collaboration, ``{Review of particle
  physics}'',} \textit{ Phys. G} \textbf{ 37} (2010) 075021,
\href{http://dx.doi.org/10.1088/0954-3899/37/7A/075021}{\doi{10.1088/0954-3899/37/7A/075021}}.
%%CITATION = JPHGB,G37,075021;%%.

\bibitem{Allanach:2002uq}
\hrefCMSnoop {} {B.~C. Allanach, ``SOFTSUSY: a program for calculating
  supersymmetric spectra'',} \textit{ Comput. Phys. Commun.} \textbf{ 143}
  (2002) 305,
  \href{http://dx.doi.org/10.1016/S0010-4655(01)00460-X}{\doi{10.1016/S0010-4655(01)00460-X}}.

\bibitem{Beenakker:1997kx}
\hrefCMSnoop {} {W.~Beenakker {et~al.}, ``Squark and Gluino Production at
  Hadron Colliders'',} \textit{ Nucl. Phys. B} \textbf{ 492} (1997) 51,
  \href{http://dx.doi.org/10.1016/S0550-3213(97)00084-9}{\doi{10.1016/S0550-3213(97)00084-9}}.

\bibitem{PDF4LHC}
\hrefCMSnoop {} {M.~Botje {et~al.}, ``{The PDF4LHC Working Group Interim
  Recommendations}'',} (2011).
  \href{http://www.arXiv.org/abs/1101.0538}{\texttt{ arXiv:1101.0538}}.

\bibitem{LEPSUSY}
\href {http://lepsusy.web.cern.ch/lepsusy/Welcome.html} {{LEPSUSYWG, ALEPH,
  DELPHI, L3 and OPAL experiments}, ``{LSP} mass limit in Minimal {SUGRA}'',}.
  LEPSUSYWG/02-06.2.

\end{thebibliography}\endgroup

\cleardoublepage \appendix\section{The CMS Collaboration \label{app:collab}}\begin{sloppypar}\hyphenpenalty=5000\widowpenalty=500\clubpenalty=5000\textbf{Yerevan Physics Institute,  Yerevan,  Armenia}\\*[0pt]
S.~Chatrchyan, V.~Khachatryan, A.M.~Sirunyan, A.~Tumasyan
\vskip\cmsinstskip
\textbf{Institut f\"{u}r Hochenergiephysik der OeAW,  Wien,  Austria}\\*[0pt]
W.~Adam, T.~Bergauer, M.~Dragicevic, J.~Er\"{o}, C.~Fabjan, M.~Friedl, R.~Fr\"{u}hwirth, V.M.~Ghete, J.~Hammer, N.~H\"{o}rmann, J.~Hrubec, M.~Jeitler, W.~Kiesenhofer, V.~Kn\"{u}nz, M.~Krammer, D.~Liko, I.~Mikulec, M.~Pernicka$^{\textrm{\dag}}$, B.~Rahbaran, C.~Rohringer, H.~Rohringer, R.~Sch\"{o}fbeck, J.~Strauss, A.~Taurok, P.~Wagner, W.~Waltenberger, G.~Walzel, E.~Widl, C.-E.~Wulz
% ...
\vskip\cmsinstskip
\dag:~Deceased\\
% ...

\end{sloppypar}
\end{document}